\documentclass[useAMS,usenatbib,letterpaper]{mn2e}
\usepackage{graphicx, lscape}
\usepackage{epsfig}
\usepackage{booktabs}
\usepackage{amsmath, amssymb}

\def\vhel{\ifmmode{V_{{\rm HEL}}}\else{$V_{{\rm HEL}}$}\fi}
\def\vsys{\ifmmode{V_{\rm sys}}\else{$V_{\rm sys}$}\fi}
\def\kms{\ifmmode{~{\rm km\,s}^{-1}}\else{~km~s$^{-1}$}\fi}
\def\vlsr{\ifmmode{v_{\rm lsr}}\else{$v_{\rm lsr}$}\fi}

\def\ltsim{\ifmmode\stackrel{<}{_{\sim}}\else$\stackrel{<}{_{\sim}}$\fi}
\def\gtsim{\ifmmode\stackrel{>}{_{\sim}}\else$\stackrel{>}{_{\sim}}$\fi}

\def\reff@jnl#1{{\rm#1\/}}

\def\aj{\reff@jnl{AJ}}                  
\def\araa{\reff@jnl{ARA\&A}}            
\def\apj{\reff@jnl{ApJ}}                
\def\apjl{\reff@jnl{ApJ}}               
\def\apjs{\reff@jnl{ApJS}}              
\def\ao{\reff@jnl{Appl.Optics}}         
\def\apss{\reff@jnl{Ap\&SS}}            
\def\aap{\reff@jnl{A\&A}}               
\def\aapr{\reff@jnl{A\&A~Rev.}}         
\def\aaps{\reff@jnl{A\&AS}}             
\def\azh{\reff@jnl{AZh}}                        
\def\baas{\reff@jnl{BAAS}}              
\def\jrasc{\reff@jnl{JRASC}}            
\def\memras{\reff@jnl{MmRAS}}           
\def\mnras{\reff@jnl{MNRAS}}            
\def\pra{\reff@jnl{Phys.Rev.A}}         
\def\prb{\reff@jnl{Phys.Rev.B}}         
\def\prc{\reff@jnl{Phys.Rev.C}}         
\def\prd{\reff@jnl{Phys.Rev.D}}         
\def\prl{\reff@jnl{Phys.Rev.Lett}}      
\def\pasp{\reff@jnl{PASP}}              
\def\pasj{\reff@jnl{PASJ}}              
\def\qjras{\reff@jnl{QJRAS}}            
\def\skytel{\reff@jnl{S\&T}}            
\def\solphys{\reff@jnl{Solar~Phys.}}    
\def\sovast{\reff@jnl{Soviet~Ast.}}     
 \def\ssr{\reff@jnl{Space~Sci.Rev.}}     
\def\zap{\reff@jnl{ZAp}}                        

\def\LaTeX{L\kern-.36em\raise.3ex\hbox{a}\kern-.15em
    T\kern-.1667em\lower.7ex\hbox{E}\kern-.125emX}

\begin{document}

\title[]{A 33 GHz VSA  survey of the Galactic plane from $\ell=27^{\circ}$ to $46^{\circ}$}
\author[M.Todorovi\'{c} et al.]{Magdolna Todorovi\'c,$\!^{1}$ 
Rodney D. Davies,$\!^{1}$ Clive Dickinson,$\!^{1}$ Richard J. Davis,$\!^{1}$
\newauthor Kieran A. Cleary,$\!^{2}$ Ricardo Genova-Santos,$\!^{3}$ Keith J.B. Grainge,$\!^{4,5}$ Yaser A. Hafez,$\!^{6}$ 
\newauthor Michael P. Hobson,$\!^{4}$  Michael E. Jones,$\!^{7}$ Katy Lancaster,$\!^{8}$ Rafael Rebolo,$\!^{3}$ 
\newauthor Wolfgang Reich,$\!^{9}$ Jos\'{e} Alberto Rubi\~{n}o-Martin,$\!^{3}$ Richard D.E. Saunders,$\!^{4,5}$ 
\newauthor Richard S. Savage,$\!^{4}$ Paul F. Scott,$\!^{4}$ An\v{z}e Slosar,$\!^{10}$ Angela C. Taylor,$\!^{7}$ 
\newauthor Robert A. Watson$^{1}$
\\
$^{1}$Jodrell Bank Centre for Astrophysics, School of Physics \& Astronomy, University of Manchester, Oxford Road, Manchester M13 9PL, U.K.\\
$^{2}$Cahill Center for Astronomy and Astrophysics, California Institute of Technology, 1200 E California Blvd,, Pasadena, CA 91125, U.S.A.\\
$^{3}$Instituto de Astrofis\'ica de Canarias, 38200 La Laguna, Tenerife, Canary Islands, Spain \\
$^{4}$Astrophysics Group, Cavendish Laboratory, University of Cambridge, 19 J.J. Thompson Avenue, Cambridge, CB3 0HE, U.K. \\
$^{5}$Kavli Institute for Cosmology, Cambridge, Madingley Road, Cambridge, CB3 0HA, U.K. \\
$^{6}$National Center for Mathematics and Physics, KACST, PO Box 6086, Riyadh 11442, Saudi Arabia \\
$^{7}$Oxford Astrophysics, University of Oxford, Denys Wilkinson Building, Keble Road, Oxford, OX1 3RH, U.K. \\
$^{8}$University of Bristol, Tyndall Avenue, Bristol BS8 ITL, U.K.\\
$^{9}$Max-Planck-Institut f\"{u}r Radioastronomie, Auf dem H\"{u}gel 69, 53121 Bonn, Germany \\
$^{10}$Brookhaven National Laboratory, Upton NY 11973, U.S.A.
}

\maketitle
\label{firstpage}

\begin{abstract}
The Very Small Array (VSA) has been used to survey the $\ell \sim27^{\circ}$ to $\sim46^{\circ}$, 
$|{\it{b}}|<4^{\circ}$ region of the Galactic plane at a resolution of 13
arcmin. This
$\ell$-range covers a section through the Local, Sagittarius and the Cetus spiral arms.  The
survey consists of 44 pointings of the VSA, each with a r.m.s. sensitivity of $\sim90$~mJy~beam$^{-1}$.  These
data are combined in a mosaic to produce a map of the area.  The majority of the sources  
within the map are HII regions. 

The main aim of the programme was to investigate the anomalous radio emission from the warm
dust
in individual HII regions of the survey.  This programme required making a spectrum extending
from GHz frequencies to the FIR IRAS frequencies for each of 9 strong sources selected to lie in
unconfused areas.  It was necessary to process each of the frequency maps with the same
{\it{u,v}} coverage as was used for the VSA 33 GHz observations.  The additional radio data
were at 1.4, 2.7, 4.85, 8.35, 10.55, 14.35 and 94 GHz in addition to the 100, 60, 25 and
12 $\mu$m IRAS bands. From each spectrum the free-free, thermal dust and anomalous dust emission
were determined for each HII region.  The mean ratio of 33 GHz anomalous flux density to
FIR 100 $\mu$m flux density for the 9 selected HII regions was $\Delta S(33 ~{\rm GHz})/S(100~\mu$m$) = 1.10 \pm
0.21 \times10^{-4}$. When combined with 6 HII regions previously observed with the VSA and the
CBI, the anomalous emission from warm dust in HII regions is detected with a 33 GHz emissivity
of $4.65 \pm 0.40$  $\mu$K  (MJy/sr)$^{-1}$ ($11.5\sigma$). This level of anomalous emission is 0.3
to 0.5 of that detected in cool dust clouds. 

A radio spectrum of the HII region anomalous emission covering GHz frequencies is constructed.
 It has the shape expected for spinning dust comprised
of very small grains.  The anomalous radio emission in HII regions  is on average $41\pm 10
$~per cent of the radio continuum at 33 GHz.  Another result is that the excess (i.e.
non-free-free) emission from HII regions at 94 GHz correlates strongly with the 100 $\mu$m
emission; it is also inversely correlated with the dust temperature.  Both these latter results
are as expected for very large grain dust emission.  The anomalous emission on the other hand
is expected to originate in very small spinning grains and correlates more closely with 
the 25 $\mu$m emission.   
\end{abstract}



\setcounter{figure}{0}

\section{INTRODUCTION}
\label{sec:introduction}

Radio surveys of the Galactic plane show the presence of small diameter sources embedded in 
a diffuse background.  These sources have been identified as supernova remnants (SNRs) and 
HII regions, with an occasional extragalactic source.  Each has a characteristic flux density spectral 
index $\alpha$ defined as $S \sim \nu^{\alpha}$; for SNRs $\alpha$ is in the range $-$0.3 to $-$0.8 with an 
average of $-$0.5 at GHz frequencies (Green 2009), while for HII regions $\alpha$ is restricted to the 
range -0.11 to -0.13 (Dickinson, Davies \& Davis 2003) depending on frequency and electron temperature.  
The diffuse emission centred about the Galactic plane has conventionally been interpreted as a mix of 
synchrotron emission with $\alpha \sim -0.7$ at $\nu< 1$ GHz and free-free emission.  Synchrotron 
dominates at $\nu <1$ GHz.   

In the last 10 years this picture has changed radically following the identification of a third emission 
component which is strongly correlated with interstellar dust as mapped at FIR wavelengths by IRAS 
for example.  This component was first identified in the early searches for fluctuations in the Cosmic 
Microwave Background (CMB) at frequencies in the range 10 to 60 GHz (Leitch et
al. 1997; de Oliveira-Costa et al., 1999, 2002, 2004; Kogut et al. 1996) when it
was variously interpreted as emission 
from very hot or from normal interstellar ionized gas which was correlated with FIR dust emission.  
The situation became clearer when all-sky radio and H$\alpha$ surveys became available along with an emission 
model for spinning dust (Draine \& Lazarian 1998) to complement the all-sky multi-frequency survey 
by COBE and later by WMAP. The three  components could then be separated and evaluated (Banday et al. 2003, 
Lagache 2003; Davies et al. 2006; Miville-Desch{\^e}nes \& Lagache 2005).  

The definitive determinations of the spectrum of spinning dust have come from the observations of compact dust 
clouds both at intermediate latitudes and near the Galactic plane.  The best examples are the Perseus molecular 
cloud (Watson et al. 2005, Tibbs et al. 2010) and the Lynds dark nebula LDN1622 (Finkbeiner et al. 2002, 2004; Casassus et al. 2006). Additional detections in dark clouds have recently been reported (Scaife et al.~2009). The 
peak in the emission is in the range 20 to 30 GHz and is at a level compatible with the Draine \& Lazarian theory.

 Many of the dust clouds studied are in star-forming regions which include HII regions; the additional free-free 
emission complicates the situation; this problem can be resolved by using data from a range of radio frequencies 
to separate the emission components.  Nonetheless, tentative detections of anomalous emission have been made 
in some bright HII regions (Dickinson et al.~2007, 2009) while no detections or upper limits have been made in others 
(Dickinson et al. 2006; Scaife et al. 2008).  The present 33 GHz survey of a rich $19^{\circ} \times 6^{\circ}$ area 
of the northern Galactic plane is aimed at significantly improving knowledge of anomalous emission from bright 
HII regions and at the same time identifying the detected sources as HII regions, SNRs or extragalactic objects.  
Anomalous emission from HII regions is predicted by spinning dust models (Draine \& Lazarian 1998,  
Ali-Ha{\"i}moud et al. 2009).

The paper is organized as follows.  Section 2 gives a summary of the parameters of the Very Small Array (VSA) when operated in 
its extended configuration.  Section 3 describes the calibration and data reduction methods.  The procedure 
for producing the maps, including the mosaicing technique as applied to both the VSA 33 GHz data and the ancillary 
data, is given in Section 4.  In Section 5 we describe the method of extracting flux densities and dimensions of the brightest sources.  Section 6 
describes the astronomical results of the survey which include a clear determination of the spectrum of spinning dust 
in HII regions and a description of the physical conditions in the warm dust associated with the detected HII regions. Section~7 discusses the properties of the excess emission while Section~8 discusses other properties of the HII regions. Conclusions are given in Section 9.


\section{THE VSA EXTENDED ARRAY}
\label{sec:extended_array}

The VSA is an interferometer situated at the high and dry site of the El Teide Observatory, Tenerife, at an 
altitude of 2340 m. The array  operates in the $\it{Ka}$ band (26-36 GHz); the observations
for this Galactic plane survey were made at a centre frequency of 33 GHz and a
bandwidth of $\Delta \nu \sim$  1.5~GHz.  The 14 horns of the array are
mounted on a $4 \times 3$ m$^{2}$ tip-tilt table, inside a metal screen which minimizes ground
spillover.  In the extended mode of the VSA each corrugated horn feeds a 322~mm aperture mirror; the 
baselines available are in the range 0.6 to 2.5~m.  Each horn feed has a primary beamwidth of FWHM 
$2^{\circ}.12$ at 33 GHz.  The effective resolution of the extended array is approximately 13 arcmin (uniform weighting) at 33 GHz.
The geometry of the tip-tilt table and the screen restrict observations to elevations above 
$\sim 55^{\circ}$;  the corresponding declination range is $-5^{\circ}$ to $+65^{\circ}$.   

The point source sensitivity of the extended VSA is $\sim 6$~Jy~s$^{1/2}$ at an average system temperature 
of 35~K.  The corresponding  brightness temperature sensitivity is $\sim15$~mK~s$^{1/2}$ in the synthesized beam 
area of $ \sim1.4 \times10^{-5}$~sr. Unlike the Cosmic Background Imager (CBI), which used a co-mounted array, the VSA 
does not require the subtraction of a lead/trail reference field. This is because the tracking VSA antennas provide a 
non-zero astronomical fringe rate. 

\begin{table*}
\centering
\caption{VSA specs.\label{tab:vsa_specs}}
\begin{tabular}{l c}
\hline
Description 				& phase-switching interferometer\\
Location 					& Iza\~{n}a, Tenerife\\
Altitude 					& 2340 m\\
Number of antennas (baselines) 	& 14 (91)\\
Baseline lengths 			& 0.6 m - 2.5 m\\
Centre frequency 			& 33 GHz\\
Bandwidth, $\Delta\nu$		& 1.5 GHz\\
Correlator 					& 91 channel complex correlator\\
Mirror size 					& 322 mm\\
Primary beam (FWHM) 		&  $2^{\circ}.12$\\
Synthesized beam (FWHM)  	& $\sim13'$\\
System temperature, $T_{sys}$ 	& $\sim35$ K\\
Point-source flux sensitivity 		& $\sim 6$ Jy s$^{1/2}$\\
Temperature sensitivity 		& $\sim 15$ mK s$^{1/2}$\\
\hline
\end{tabular}
\end{table*}

Table~\ref{tab:vsa_specs} summarizes the specification of the extended VSA. 
Further details of the technical specifications of the VSA may be found in Watson et al. (2003), Scott et al. (2003) and 
Maisinger et al. (2003).


\section{OBSERVATIONS AND DATA REDUCTION}
\label{sec:abs_and_data}


\subsection{Observations}
\label{sec:obs_schedule}

\begin{figure*}
\begin{center}
\caption{The solid circles show the pointing positions
for the 44 VSA fields superposed on the 33-GHz VSA mosaic. The size of the circles
is the FWHM primary beam of the VSA in extended mode.\label{fig:circles}}
\includegraphics[width=9cm]{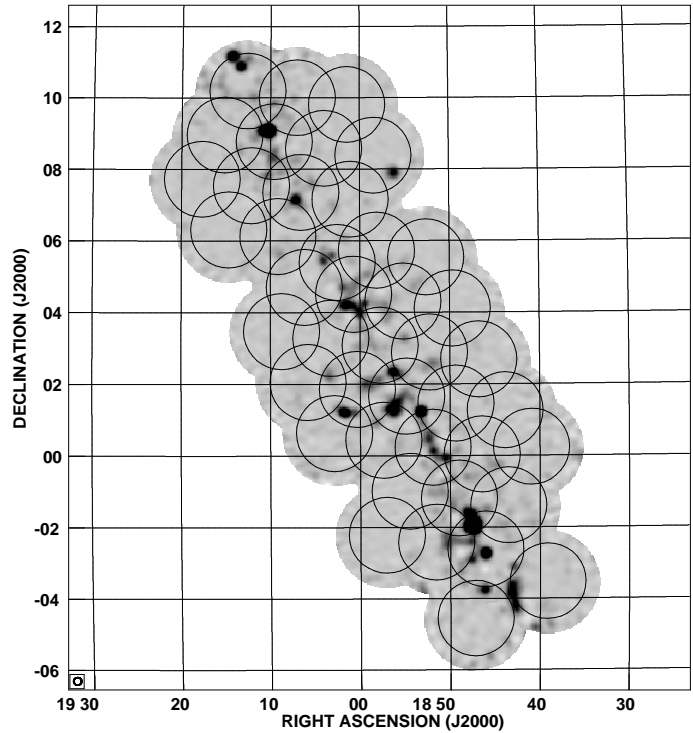}
\end{center}
\end{figure*}

The 33-GHz Galactic plane survey was made between July 2002 and August 2003, when the  
VSA was in the extended configuration. Fig. \ref{fig:circles} shows the area of the Galactic plane
covered which extends northwards from the southern limit of the VSA (Dec $= -5^{\circ}$). The circles show 
half-power beamwidths of the 44 fields observed. The outer envelope of the survey represents the $\frac{1}{4}$-power 
level of the survey; a map can be reliably reconstructed within this 114 deg$^{2}$ area. The overlap of the fields 
ensures that more than one independent observation is made of each source in the survey area.  
Three fields are missing for complete coverage of the area (G332$-$226,
G396$-$226, G396$+$226).

On any day  the VSA observed one field for typically $\sim2.4$ hours. Some of the low declination fields were 
observed more than once. Calibrator observations of nearby  
bright radio sources were interleaved with the field observations. These were
typically observed for a minimum of 10 minutes. The properties of the calibrators are discussed in detail 
in Sections~\ref{sec:calibration} and ~\ref{sec:abs_cal}.

The  44 pointings are listed in Table~\ref{tab:fieldpos_extended}. Note that the actual integration time is typically $\sim80$~per cent 
of the values indicated in the table, due to data editing and atmospheric effects, as discussed in the following section.



\begin{table}
\centering
 \caption{The Galactic coordinates  of the  centres of fields observed with the extended
VSA. Column 4 lists the integration time for each field.} \label{tab:fieldpos_extended} 
  \begin{tabular}{lccc}
    \hline
Field    & $\ell$ & $\emph{b}$ & $t_{\rm int}$ (hrs) \\
       \hline
G284-113&	28.4& 		-1.13&      		5.03\\
G284+113&	28.4&		+1.13&		2.43\\
G300+000&	30.0&		0.00&		7.53\\
G308-113&	30.8&		-1.13&		7.39\\
G308+113&	30.8&		+1.13& 		2.46\\
G316+000&	31.6&		0.00&		2.46\\
G316-226&	31.6&		-2.26&		3.21\\
G316+226&	31.6& 		+2.26&		2.45\\
G324-113&	32.4&		-1.13&		2.28\\
G324+113&	32.4&		+1.13&		2.46\\
G332+000&	33.2&		0.00&		2.31\\
G332-226&	33.2&		-2.26&		2.30\\
G332+226&	33.2&		+2.26&		2.18\\
G340-113&	34.0&		-1.13&		2.41\\
G340+113&	34.0&		+1.13&		2.32\\
G348+000&	34.8&		0.00&		2.32\\
G348-226&	34.8&		-2.26&		2.47\\
G348+226&	34.8&		+2.26&		2.31\\
G356-113&	35.6&		-1.13&		2.41\\
G356+113&	35.6&		+1.13&		2.42\\
G364+000&	36.4&		0.00&		2.40\\
G364-226&	36.4&		-2.26&		2.39\\
G364+226&	36.4&		+2.26&		2.32\\
G372-113&	37.2&		-1.13&		2.32\\
G372+113&	37.2&		+1.13&		2.34\\
G380+000&	38.0&		0.00&		2.31\\
G380-226&        38.0&		-2.26&		2.31\\
G380+226&       38.0&		+2.26&		2.34\\
G388-113&	38.8&		-1.13&		2.29\\
G388+113&	38.8&		+1.13&		2.31\\
G396+000&	39.6&		0.00&		2.45\\
G404-113&	40.4&		-1.13&		2.21\\
G404+113&	40.4&		+1.13&		2.25\\
G412+000&	41.2&		0.00&		2.37\\
G412-226&   	41.2&		-2.26&		2.42\\
G412+226&   	41.2&		+2.26&		2.43\\
G420-113&	42.0&		-1.13&		2.51\\
G420+113&	42.0&		+1.13&		2.31\\
G428+000&	42.8&		0.00&		2.38\\
G428-226&    	42.8&		-2.26&		2.37\\
G428+226&   	42.8&		+2.26&		2.38\\
G436-113&	43.6&		-1.13&		2.37\\
G436+113&	43.6&		+1.13&		2.34\\
G444+000&	44.4&		0.00&		2.35\\
  \hline
 \end{tabular}
\end{table}


\subsection{Data reduction pipeline}
\label{sec:data_reduction}

The data reduction and calibration of all the 44 fields were made with the REDUCE  package 
which was written specifically for the VSA (e.g. Dickinson et al. 2004). Every field was analyzed separately.
The data for each field were first checked  for atmospheric, pointing and geometry errors;
data lying outside the allowed parameter range were automatically flagged. Gain and phase corrections based on the calibration 
observations were then implemented. 

 Fourier filtering was used to remove correlated signals due to ground effects and the Sun or Moon if nearby (Watson et al., 2003). 
This filtering process has no effect on the data because of the difference in fringe rates between the astronomical and non-astronomical signals. 
 Data were filtered if the Sun and the Moon were within 27$^{\circ}$ and 18$^{\circ}$ of the field centres, respectively. Observations 
were not used if the Sun or the Moon were within 9$^{\circ}$ of the field centre. The final step in the reduction pipeline was to take 
account of the apparent noise levels in the 91 baselines arising from the different receiver temperatures and bandwidths; the 
measured r.m.s. noise on each baseline was used to reweight the data to give optimal sensitivity.


\subsection{The daily amplitude and phase measurement}
\label{sec:calibration}

The daily observation of the Galactic plane was interleaved with observations of calibration sources whose 
positions and flux densities were known. These observations were used to calibrate the survey data in amplitude 
and phase. The phases of each of the 91 baselines could be estimated to better than $\sim10^{\circ}$.
 The overall system noise 
was continuously compared using noise signals injected into each of the 14 feed horns; variations in the system noise 
in each receiver due to changing atmospheric emission were used to correct the astronomical amplitudes. 
These corrections were typically less than a few percent. Any data with a correction factor of 
more than $10$~per cent were discarded.


\subsection{Absolute flux calibration}
\label{sec:abs_cal}

The flux densities and the brightness temperatures used here are the 33 GHz values determined at epoch 2001 from VSA 
measurements in the period 2001 - 2004 (Hafez et al, 2008). The ``absolute" calibration was made in terms of 
the brightness temperature of the planet Jupiter given by the WMAP 5-year data (Hill et al. 2009), namely 
$T_{b}=146.6\pm0.75$ K.  $T_{b}$ values for the planets were converted to flux densities using the Ephemeris 
values of the planet areas.  The radio sources Cas A and Tau A are sufficiently extended ($\sim5$-arcmin diameter) 
that corrections were required to the data, as given by Hafez et al. (2008). Over this period the sources Cas A, 
Tau A and NGC7027 were decreasing by several tenths of 1 percent per annum at 33 GHz.

The flux density scale provided by the calibrators is believed to be better than $2$~per cent; it is tied to the Jupiter 
brightness temperature as described above.

\section{IMAGING AND MOSAICING THE 33-GHz DATA}
\label{sec:mosaicing}

We describe here the main steps in imaging the individual VSA fields and then combining them into a mosaic of the 
Galactic plane field.

\subsection{The individual VSA fields}
\label{sec:The individual VSA fields}

Each of the fields was imaged by applying the  DIFMAP package (Shepherd, 1997) to the visibility data, using uniform weighting to give 
optimal resolution and retain extended emission. The CLEAN algorithm (H\"ogbom, 1974) was applied to the dirty
images; typically 5000 iterations with a low gain loop of 0.01 were used. 
The resultant r.m.s. noise on a typical field 
with a 2 hour integration was $\sim90$~mJy~beam$^{-1}$. This corresponds to 7.5~Jy~beam$^{-1}$ $s^{1/2}$ which is close to the 
noise expected from $T_{sys}$ and the 1.5 GHz bandwidth (Table~\ref{tab:vsa_specs}).

\subsection{Mosaicing - the final map}
\label{sec:Mosaicing - the final map}

The individual field images were combined into a  map using the HGEOM and  LTESS tasks of the $\mathcal{AIPS}$ 
package. The  LTESS task also corrects the individual field maps for the $2^{\circ}.1$ FWHM primary beam response 
and restricts the field to the $25$~per cent beam level. The half-power and quarter-power levels are shown in Fig.~\ref{fig:circles}.

\begin{landscape}
\begin{figure}
\begin{center}
  \includegraphics[width=18.0cm]{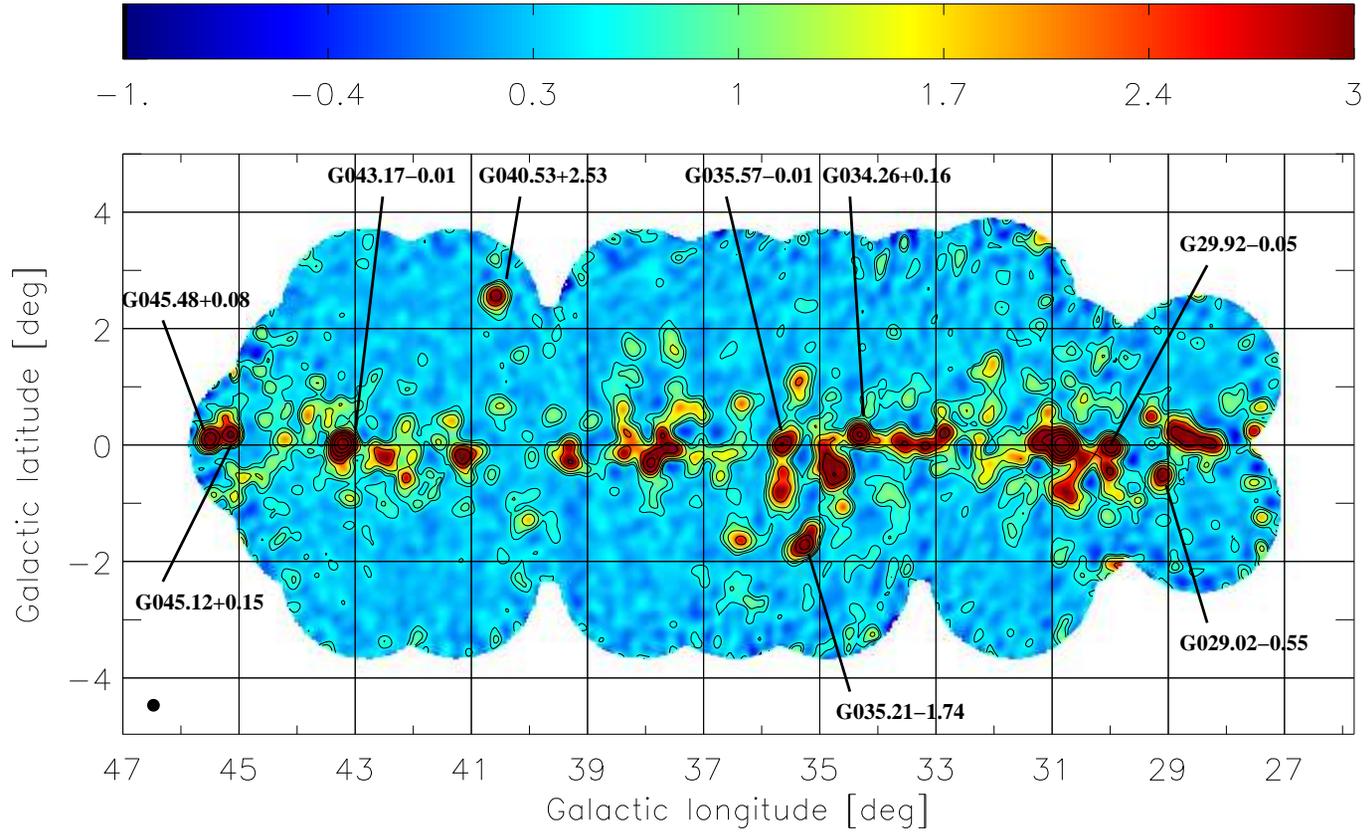}
\caption{The 33 GHz map of the Galactic plane covering the region $\ell=27^{\circ}\!$ to $46^{\circ}\!$ and ${b}=-3^{\circ}.5$ to ${b}=+3^{\circ}.5$. 
Units are Jy~beam$^{-1}$. Contour levels are at 0.5, 1, 2, 4, 8, 16, 32 and 64 per cent of the peak brightness (66.8~Jy~beam$^{-1}$).  
The synthesized beam is shown in the lower left corner.}
\label{m33}
\end{center}
\end{figure}
\end{landscape}

The final map which combines the 44 fields is shown in Fig.~\ref{m33}, plotted in Galactic coordinates. The region 
covered is $\ell=27^{\circ}$ to $46^{\circ}$ and ${b}=-3^{\circ}\!.5$ to $+3.5^{\circ}$. The areas around the edge of the map 
 have higher noise level due to the correction for beam taper. 

\subsection{The noise level on the 33-GHz map}
\label{sec:The noise level on the 33-GHz map}

The noise on the mosaiced map was estimated from the higher latitude parts of the field which were free of obvious sources. 
The effective integration time over the map is $\sim2$ hours when account is taken of overlapping fields and of the $\sim20$~per cent 
data loss due to weather, ground effects on short baselines and equipment malfunction.
The calculated average r.m.s. noise was 94~mJy~beam$^{-1}$. This value is as expected from the single field measurements.

Effective noise levels are however greater near the Galactic plane where there are strong sources as well as a diffuse background, 
$\sim 2^{\circ}$ wide  in latitude. Here the noise level may amount to $\pm1$~per cent of the adjacent source due to either limited 
coverage of the $\it{u,v}$ plane or flux leakage from incorrect phase calibration.

\section{Derivation of source parameters}
\label{sec:The 33 GHz catalogue - derivation of source parameters}

We outline here the procedures applied to derive the source parameters - flux density, position and angular size - detected in 
the survey.

\subsection{Applying JMFIT}
\label{sec:Applying JMFIT}

The task JMFIT in  $\mathcal{AIPS}$ was used to determine the parameters of the strongest and most clearly defined  
sources in the survey area. JMFIT finds the best elliptical Gaussian fit plus a baseline offset to a source and gives the peak flux density, 
the integrated flux density, the position coordinates, angular size and position angle, each with an associated error. 

Where the source is complex, JMFIT can in principle fit several components; with the S/N available in the present survey we 
restricted  the fit to 2 Gaussian components for the brightest objects. A test of the uncertainty of the source parameters was to make solutions with different 
box sizes which allowed a range of base levels to be applied.
\nopagebreak

\subsection{Confirmation with IRING}
\label{sec:Confirmation with IRING}

The  $\mathcal{AIPS}$ task IRING was used to confirm the results obtained with JMFIT, particularly where a Gaussian assumption 
was inappropriate in more complex sources. IRING requires the specification of the source centre and integrates the flux density in 
successive rings around this centre.

This procedure enables integrated flux densities to be estimated out to specified background 
levels.

The derived values of integrated flux density in most cases agreed to better than $5-10$~per cent with the values determined from JMFIT. 
Since the IRING task does not give angular size or position information, the JMFIT data are used for these parameters.

\subsection{Accuracy of the 33-GHz data}
\label{sec:Accuracy of the 33-GHz data}

As shown in Section~\ref{sec:The noise level on the 33-GHz map} the r.m.s. noise on individual fields is typically 90~mJy~beam$^{-1}$ at the 
beam centre. However, in the vicinity of the brighter sources and close to the Galactic ridge the uncertainty in flux density may be 
as much as $\sim 20$~per cent for weaker sources. In the analysis presented here, we have chosen the brightest sources detected in the VSA mosaic image and have not considered sources that are significantly confused.
Source positions are found to be accurate to $\sim1$~arcmin, from a  comparison with surveys at other frequencies, as shown in 
Section~\ref{sec:Radio positions relative to the dust}. Actual source sizes are given after deconvolution by the 13~arcmin VSA beam.

\fontsize{8}{10pt}\selectfont
\begin{table*}
\begin{center}
\begin{tabular}{c c c  l  }
\hline
Frequency [GHz] & Telescope/Survey & Angular resolution [arcmin]  & Reference \\
\hline
1.4 & Effelsberg 100 m & 9.4 & Reich et al. (1990a) \\
2.7 & Effelsberg 100 m & 4.3 & Reich et al. (1990b) \\
4.85 & Green Bank 91 m & 3.5 & Condon et al. (1994) \\
8.35 & Green Bank 13.7 m & 9.4 & Langston et al. (2000)\\
10.55 & Nobeyama 45 m & 3.0 & Handa et al. (1987)\\
14.35 & Green Bank 13.7 m & 6.6 & Langston et al. (2000)\\
94 & WMAP & 12.6 & Hinshaw et al. (2009)\\
2997 (100 $\mu$m) & IRAS & 4.3 &Miville-Desch{\^e}nes \& Lagache (2005)\\
4995 (60 $\mu$m) & IRAS & 4.0 &Miville-Desch{\^e}nes \& Lagache (2005) \\
11990 (25 $\mu$m) & IRAS & 3.8 &Miville-Desch{\^e}nes \& Lagache (2005) \\
24980  (12 $\mu$m) & IRAS & 3.8 &Miville-Desch{\^e}nes \& Lagache (2005)\\ 
\hline
\end{tabular}
\caption[]{The ancillary data used in this work.\label{ancillary_surveys}}
\end{center}
\end{table*}
\normalsize

\begin{figure*}
\begin{center}
\includegraphics[width=5.1cm]{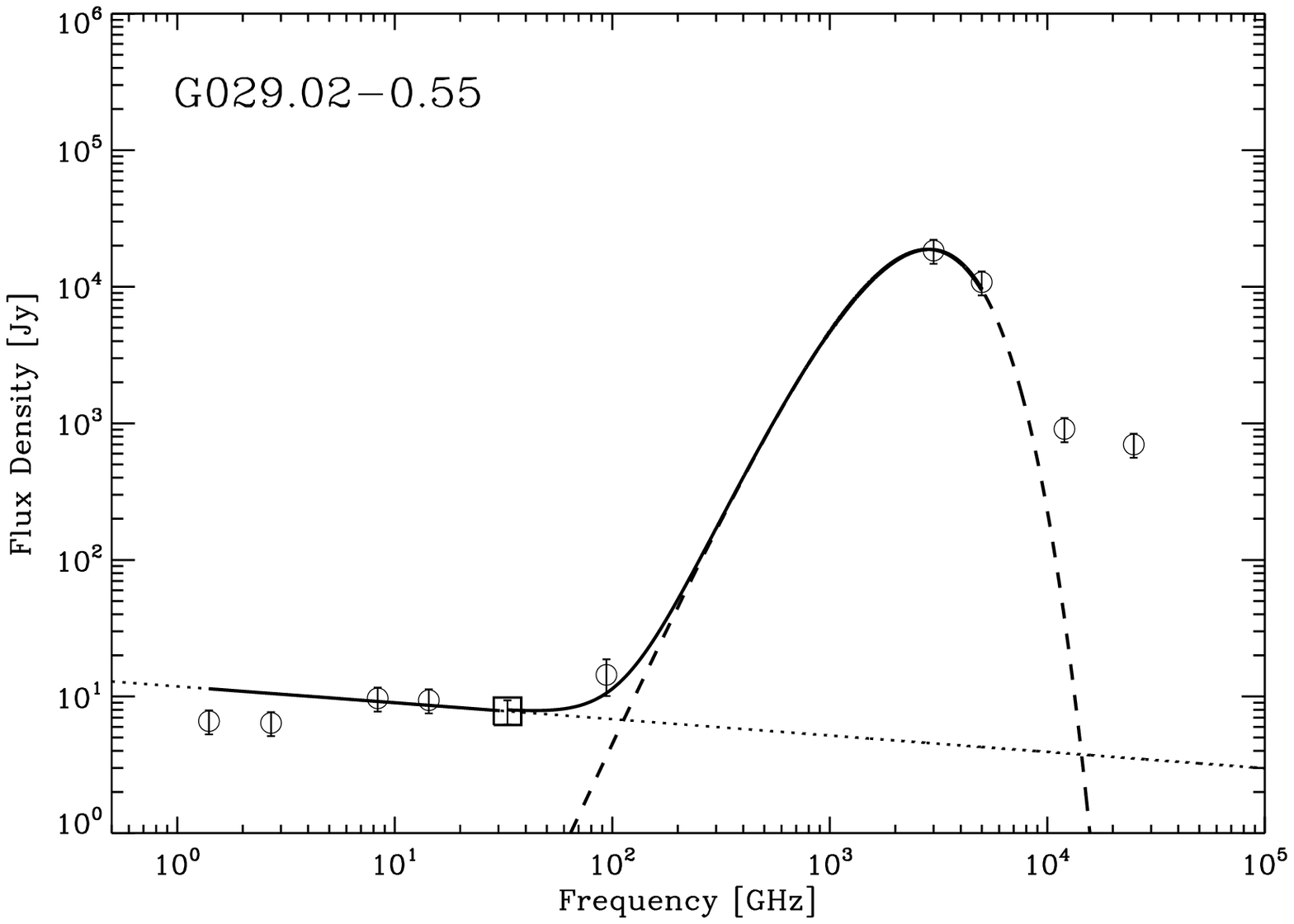}		
\includegraphics[width=5.1cm]{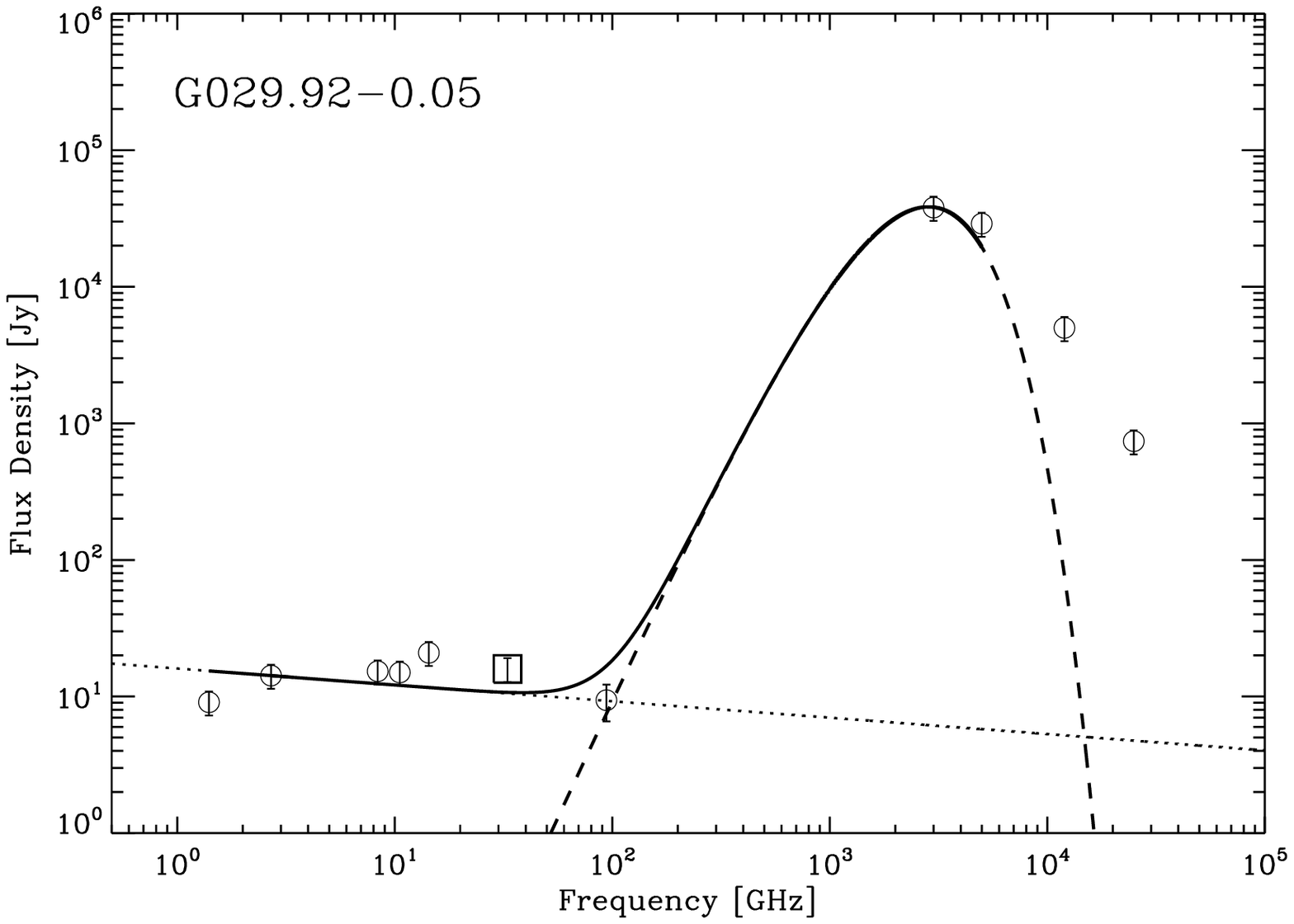}
\includegraphics[width=5.1cm]{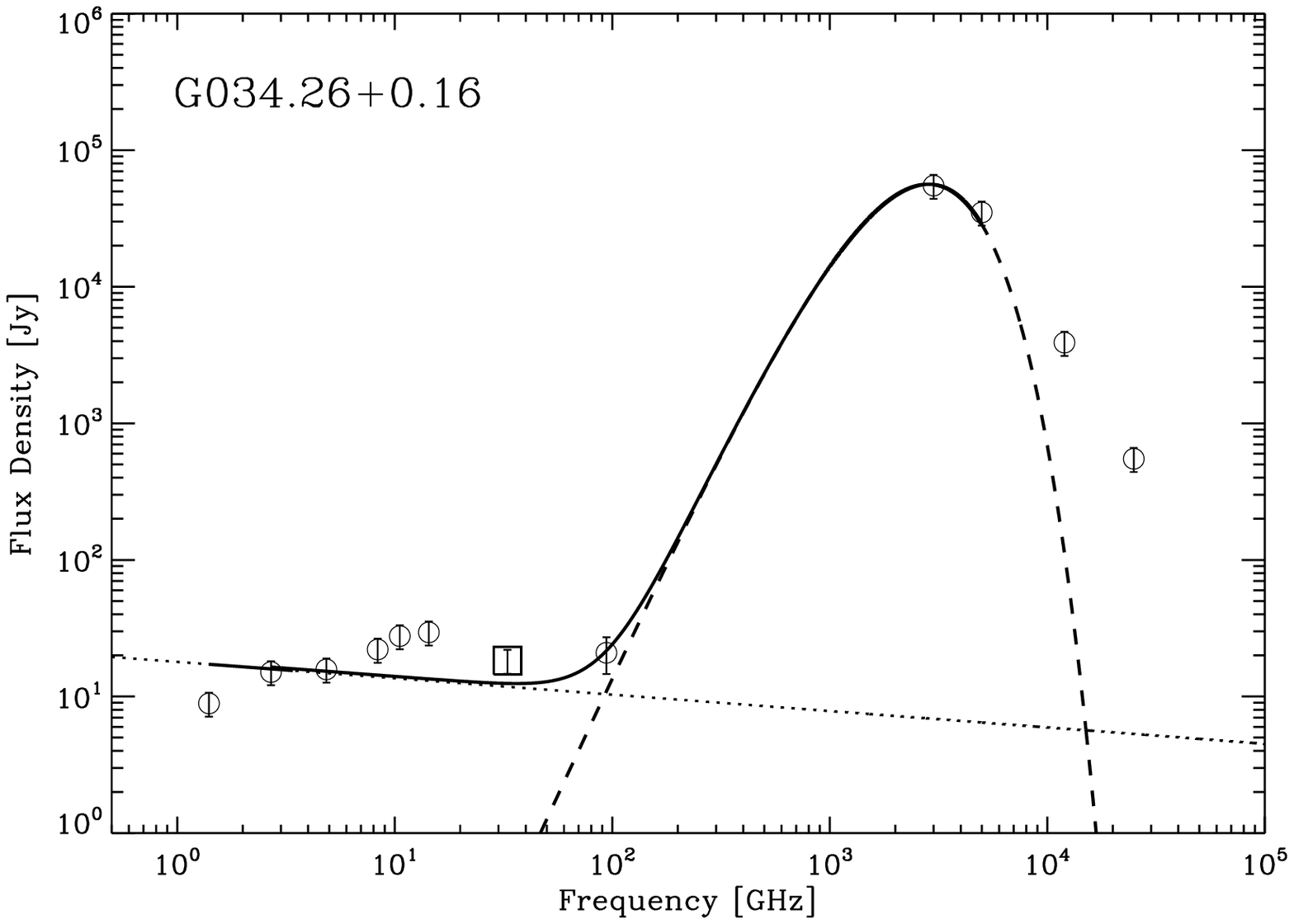}
\includegraphics[width=5.1cm]{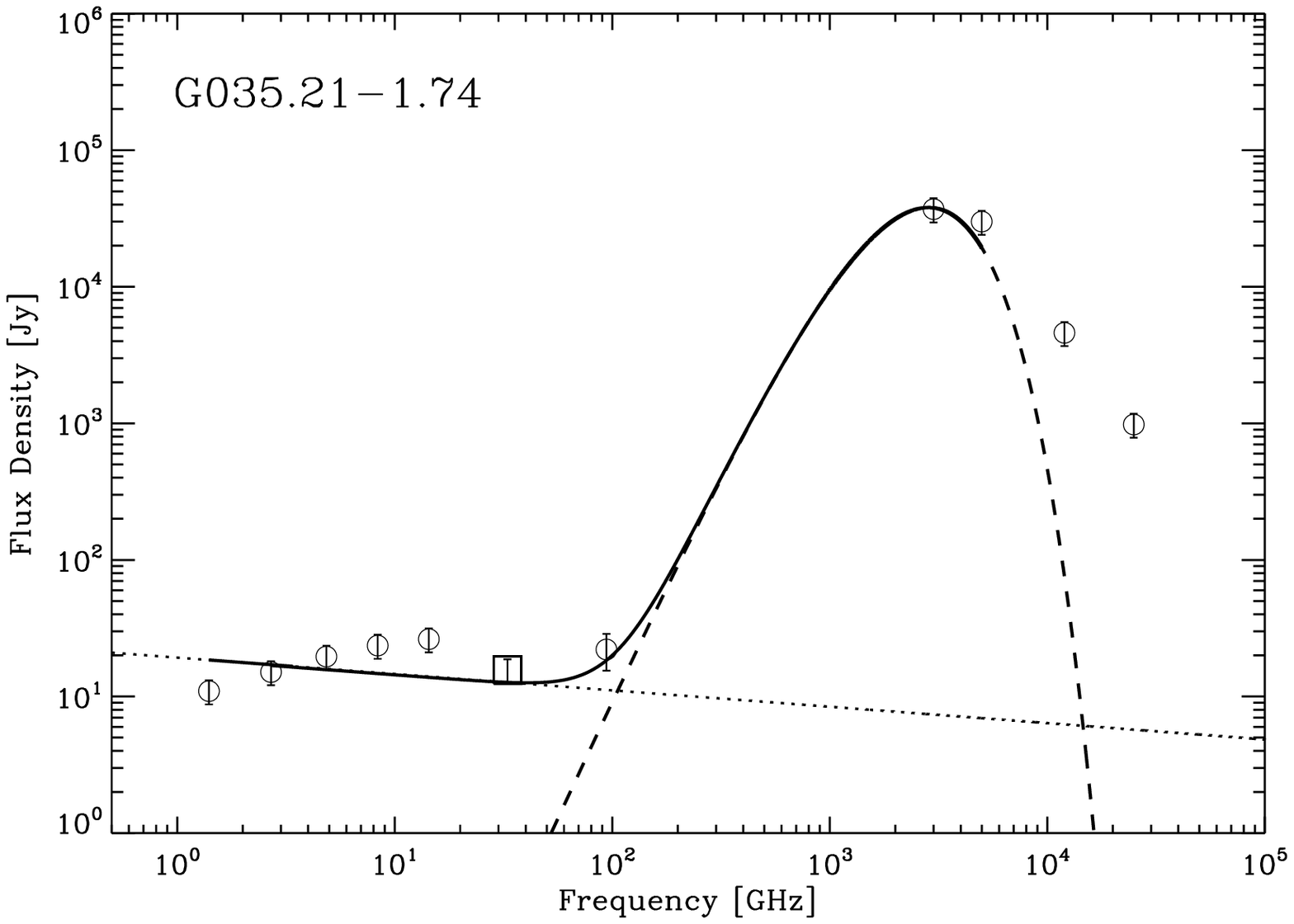}
\includegraphics[width=5.1cm]{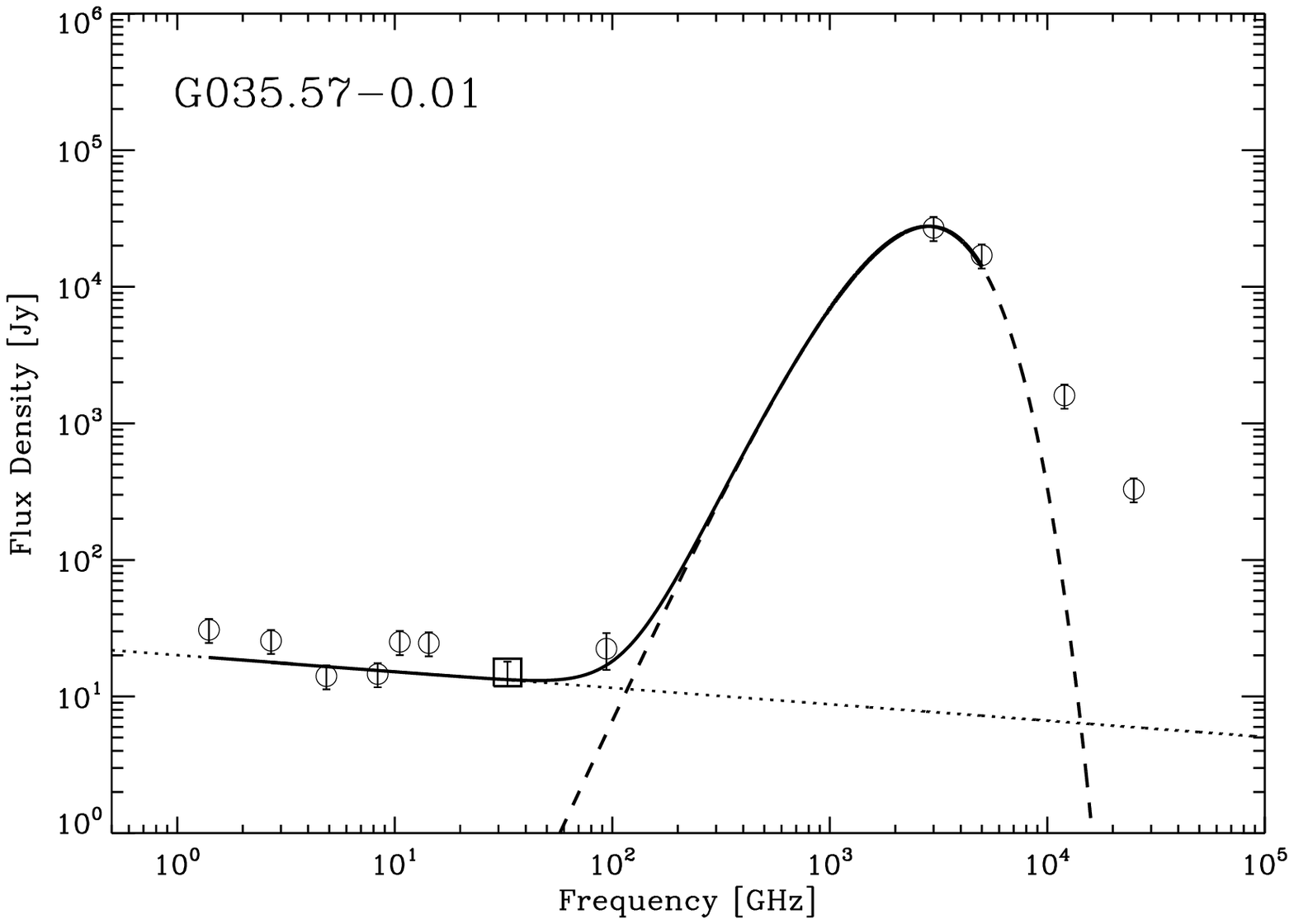}
\includegraphics[width=5.1cm]{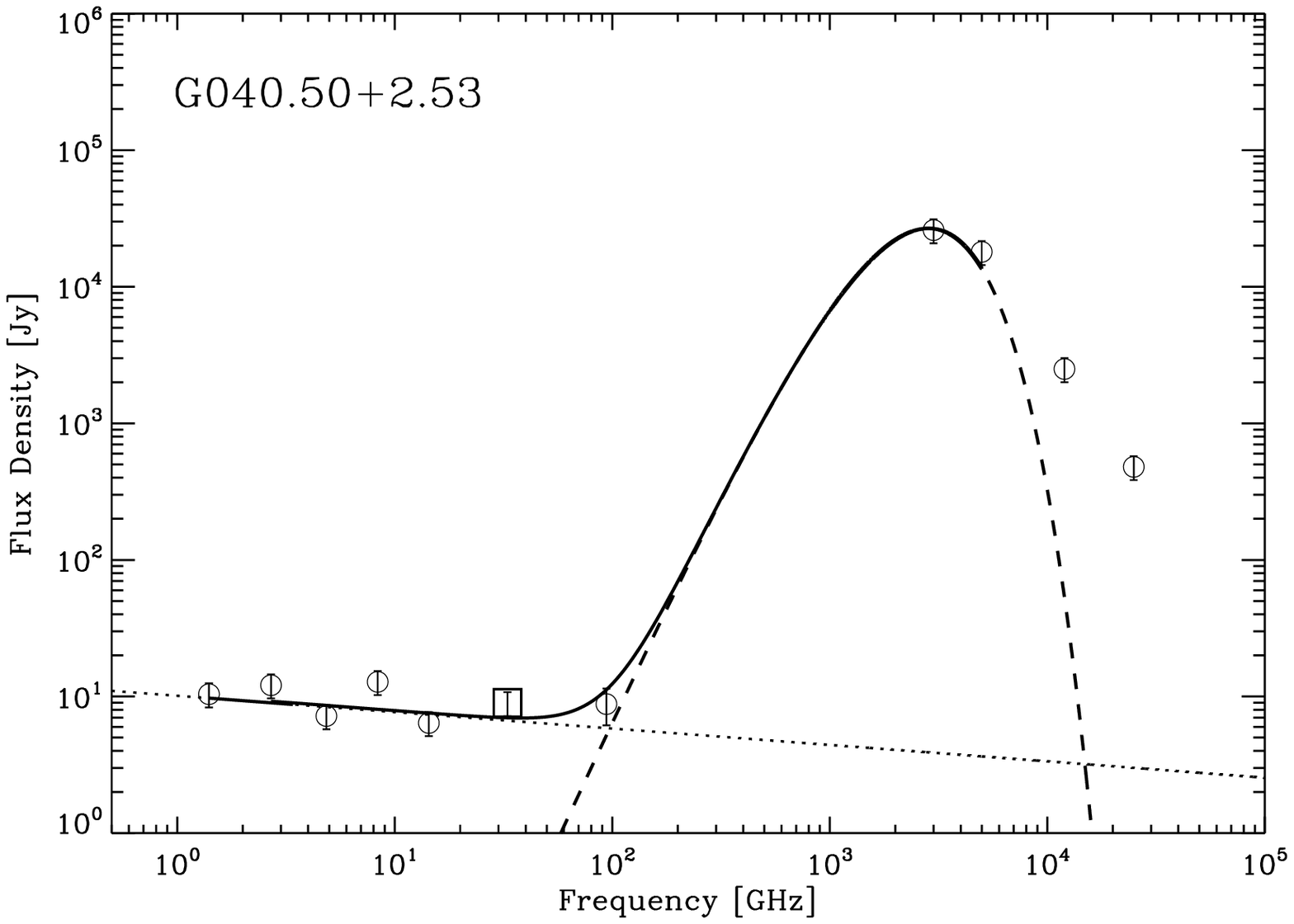}
\includegraphics[width=5.1cm]{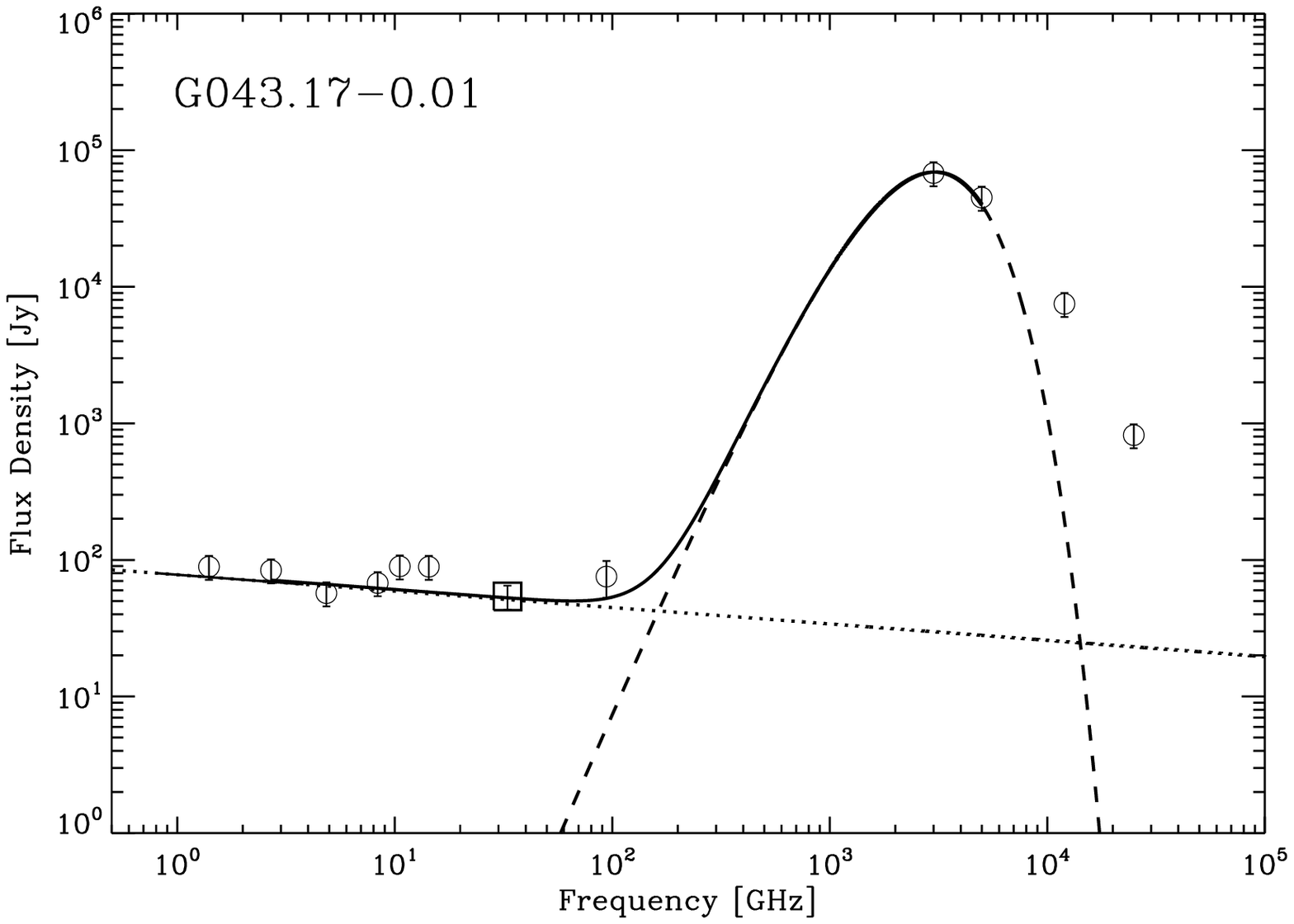}
\includegraphics[width=5.1cm]{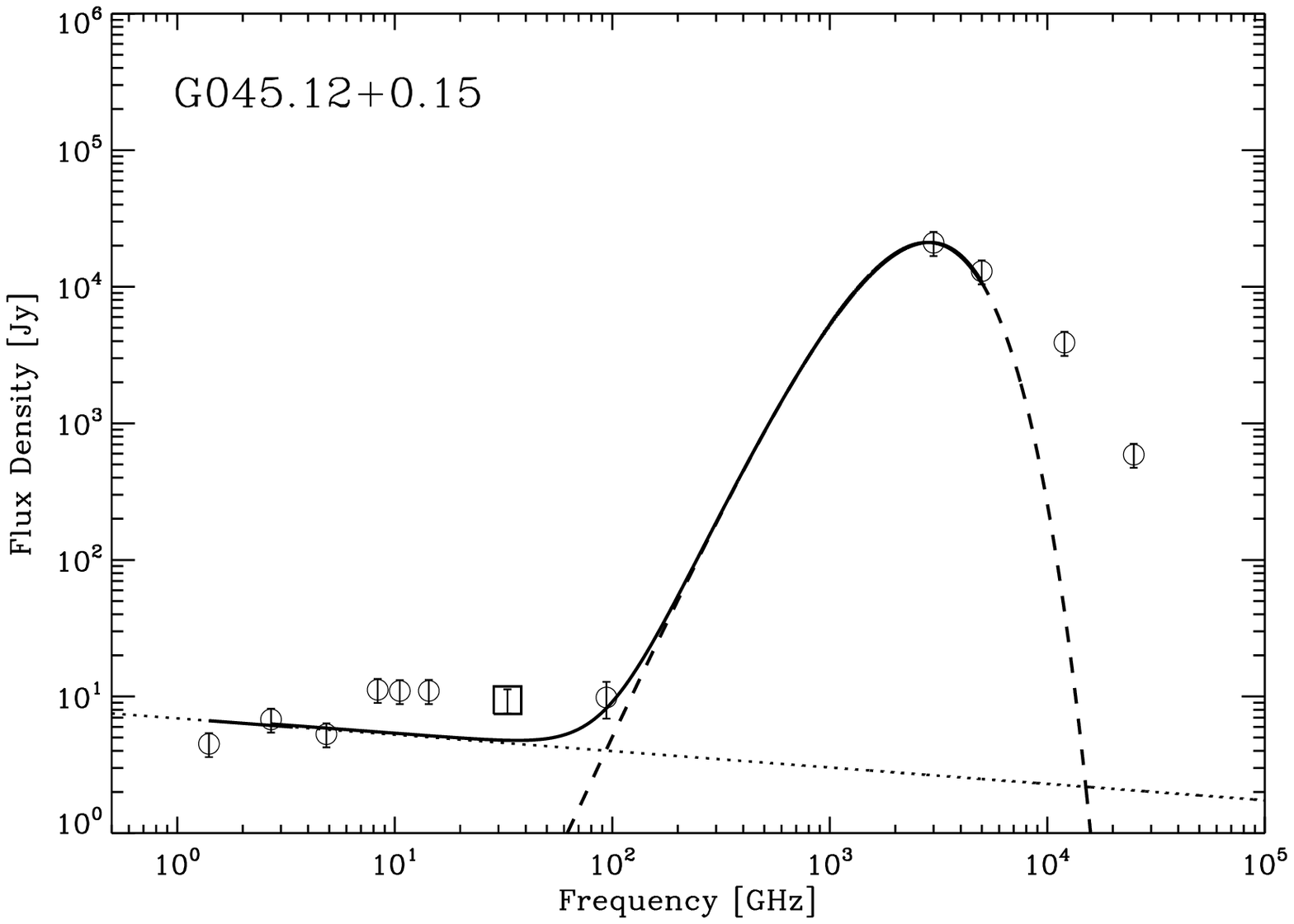}
\includegraphics[width=5.1cm]{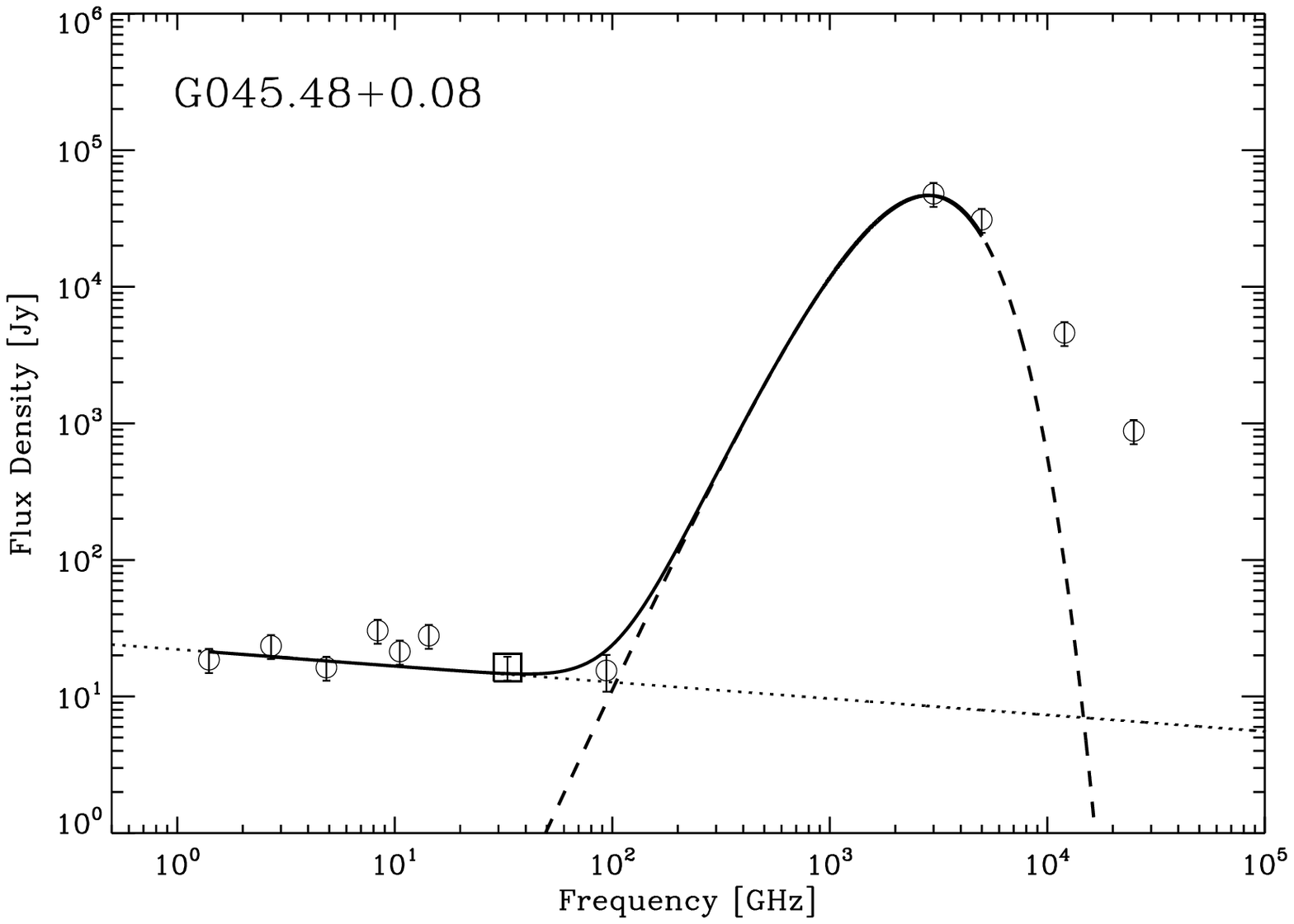}
\caption[]{Spectra of the 9 HII regions.The best-fit free-free spectrum through 
the 2.7 and 4.85 GHz data is shown (dotted line) for each 
region along with a 30 K thermal dust spectrum (dashed line) through the 100- $\mu$m and 60-$\mu$m data with $\beta =1.5$.}
\label{nine_spectra}
\end{center}
\end{figure*}

\section{SPECTRA OF SELECTED HII REGIONS}
\label{sec:spectra}

We now consider the spectral characteristics of a selection of stronger objects in the 33 GHz 
survey. A wide selection of ancillary data ranging from 1.4 GHz to the FIR, 
is available for this purpose.

\subsection{The ancillary data}
\label{sec:The ancillary data}

For accurate spectral determination we require data from high signal-to-noise surveys covering the Galactic
plane. Angular resolutions comparable to that of the VSA (13~arcmin) or better are required. The surveys  
satisfying these criteria  are at 1.4, 2.7, 4.85, 8.35, 10.55, 14.35, 33, 94~GHz and the FIR bands at 100, 60, 25 and 
12~$\mu$m and are listed in Table~\ref{ancillary_surveys}.
It can be seen that high resolution data with resolutions $\sim4-5$~arcmin are available at both radio and FIR 
wavelengths; these are particularly useful in the evaluation of the physical properties of the sources.

\subsection{Modelling and mosaicing the ancillary data}
\label{sec:Modelling and mosaicing the ancillary data}

In order to derive the spectrum of a source in the 33-GHz survey it is essential to compare the maps at different 
frequencies with the same resolution as the VSA observation. This is achieved by applying the same VSA $\it{u,v}$ 
coverage for each field, accounting for any flagged data. The procedure is undertaken in $\mathcal{AIPS}$ with 
the UVSUB task, followed by the imaging task IMAGR. When all the fields at each frequency have been modelled 
and imaged in this manner, they are  mosaiced as for the 33-GHz data to provide a map at that frequency.

\subsection{Selection of HII regions for detailed study}
\label{sec:Selection of HII regions for detailed study}

Some 60 sources can be identified in the 33-GHz map produced in the present investigation with peak 
flux densities $\geq 1.5$~Jy~beam$^{-1}$.  From this list we select the brightest objects in unconfused 
areas which are not contaminated by synchrotron sources, namely SNRs and extragalactic objects. The identification 
of the free-free and synchrotron sources is made on the evidence from the 2.7 GHz and 100 $\mu$m maps at the 
original resolution ($\sim 4'$). The Green (2009) catalogue of SNRs was  useful in this regard.

Several of the brightest features on the 33-GHz map such as G030.70-0.03 and G034.70-0.40 (the W44 region) 
had to be excluded because of their complexity and large extent; structures on scales $> 40$~arcmin are affected by 
significant flux loss and are difficult to image due to limited {\it u,v} coverage. Furthermore, the differences between the 
diffuse synchrotron and free-free 
background distributions makes the determination of a spectral index difficult, even when the same (VSA) baselines 
are applied to the ancillary data.  

Table~\ref{fitted_measured} lists the 9 HII regions with flux densities that are determined with confidence. Integrated 
flux densities are given for the 8 radio frequencies used to derive the spectra of these HII regions, namely 1.4, 2.7, 
4.85, 8.35, 10.55, 14.35, 33 and 94 GHz. The 33 GHz flux densities are all greater than 5~Jy. We note that the 
flux densities given for the HII component, A, of W49 could be derived accurately at the VSA resolution since it lies 
12~arcmin from the SNR component, B.

\subsection{Spectra of individual HII regions}
\label{sec:Spectra of individual HII regions}

The spectra of the 9 HII regions ranging from the radio to the FIR are plotted in Fig.~\ref{nine_spectra}. 
The sources of the radio and FIR 
 data are given in Table~\ref{fitted_measured}. These data are convolved to the resolution 
of the VSA data as described in Section~\ref{sec:Modelling and mosaicing the ancillary data}. 
The errors plotted for the individual observations are a conservative $\pm20$~per cent of the flux density 
(Section~\ref{sec:Accuracy of the 33-GHz data}). This value is appropriate given the observed 
scatter of data points, as described in Section~\ref{sec:Determining the free-free component}.

These spectra will now be used to derive the free-free, thermal (large grains) dust and anomalous components 
of the emission in each HII region. The anomalous component will be assessed as an excess above the other two components 
 plotted in Fig.~\ref{nine_spectra}.

\subsection{Determining the free-free component}
\label{sec:Determining the free-free component}

We use the lowest frequencies as reliable estimators of the free-free emission. Over the frequency range 1.4 to 33 GHz 
the flux density spectral index  has a well-determined value of $\beta=-0.12$ (Dickinson et al., 2003). 
The data at 2.7 and 4.85 GHz are from well-calibrated surveys, with high  ($\sim 4$~arcmin) resolution and hence provide 
the best estimators of the free-free emission. These frequencies are uncontaminated by dust (anomalous or thermal) emission.

A good test for the accuracy of the 2.7 and 4.85~GHz data, including any systematics in the analysis method, is to derive the 
average of the spectra normalized to the combined flux density at 2.7 and 4.85 GHz. The spectral index between 
2.7 and 4.85 GHz is taken into account.

\begin{figure}
\begin{center}
\includegraphics[width=8.5cm]{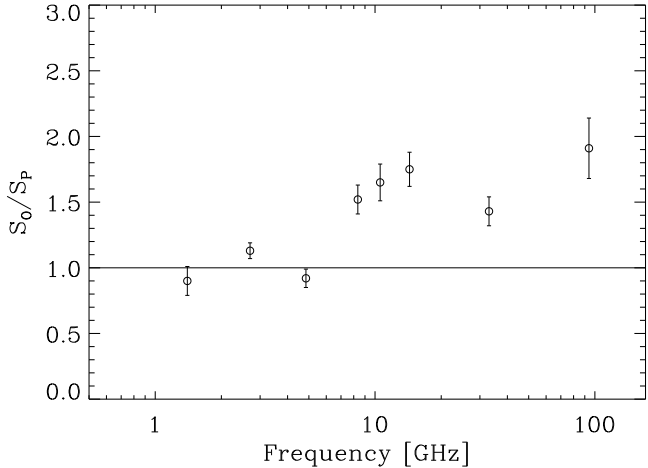}
\caption[]{The ratio of observed integrated flux density, $S_{O}$,  
to that of the predicted free-free, $S_{P}$,  for the 8 frequencies. Unity is the value for the free-free alone, based on 2.7 and 4.85 
GHz mean flux densities. $1\sigma$ errors are shown.}
\label{obs-to-free-free}
\end{center}
\end{figure}

Fig.~\ref{obs-to-free-free} shows this normalized spectrum for the 9 HII regions. The r.m.s. in the normalized flux density is shown as 
$\pm 1 \sigma$. At 2.7 and 4.85~GHz the value of the $\sigma$ of the mean is less than 6~per cent while at the higher frequencies it is $7-11$~per cent, possibly representing 
the variation in the dust contribution between HII regions.

It can be seen that the mean 1.4~GHz flux density is slightly low ($\sim 0.90$)  compared with the expected value of 1.00. 
This may represent a small degree of free-free self absorption. An HII region with a brightness temperature of 
$10^{3}$~K, and an electron temperature of 7000~K, would have a 1.4~GHz free-free optical depth of $\tau\simeq 0.15$ and a flux density of 0.85 of the optically thin value. The only HII region in this study exceeding 
this value is G045.12+0.14 which has $\tau=0.39$,  
consistent with dimensions of $1.5\times0.5$~arcmin$^{2}$, found in the high resolution NVSS 1.4~GHz survey. At 2.7 and 4.85~GHz this 
source would have $\tau=0.08$ and $\tau=0.02$ respectively. On averaging with other HII regions the effect of 
self-absorption will be no more than $10$~per cent at 1.4~GHz and a few percent at 2.7  and 4.85~GHz. The agreement of the 
1.4~GHz value with the normalized values confirms the accuracy of the calibration procedure adopted here.

\begin{landscape}
\begin{table*}
\fontsize{8}{10pt}\selectfont
\renewcommand{\arraystretch}{0.9}

\vspace{0.5cm}
\begin{center}
\caption[]{The predicted and 
observed  integrated flux densities in Jy for the 9 sources.  The letters O and P stand for 
``observed'' and ``predicted'' integrated flux density, respectively. The predicted free-free flux density is estimated from the 
2.7 and 4.85 GHz values. The excess, E, is the difference between the
predicted and observed flux densities.}
\begin{tabular}{l c r r r r r r r r r|}
\hline
Source &   & 1.4  GHz& 2.7  GHz& 4.85  GHz&  8.35  GHz&  10.55 GHz& 14.35  GHz& 33  GHz& 94  GHz \\
\hline 
G029.02$-0.55$ & O    & 6.6     & 6.4    & -   & 9.7    &   -  & 9.4   & 7.8   & 14.4\\
	      		& P    & 6.9      & 6.4    & -   & 5.6    &   -  & 5.2   & 4.7   & 4.2\\
& $\textbf{E}$    & $\textbf{-0.3}$ & $\textbf{0.0}$ & -  & $\textbf{+4.1}$ &-& $\textbf{+4.2}$ & $\textbf{+3.1}$ &$\textbf{+10.2}$\\
\hline 
G029.92$-0.05$ & O    & 9.1     &14.8   & -   & 15.2  &  15.0  & 20.9 & 15.9 & 9.4\\
	                & P     &15.4    &14.8   & -   & 12.4   & 12.1  & 11.7  & 10.5 & 9.3\\
& $\textbf{E}$    & $\textbf{-6.3}$ & $\textbf{0.0}$  & - 	& $\textbf{+2.8}$ & $\textbf{+2.9}$ & $\textbf{+9.2}$ & $\textbf{+5.4}$ & $\textbf{+0.1}$\\
\hline 
G034.26+0.16    & O    & 8.9 	& 15.1  & 15.8 	& 22.1 & 27.7 	& 29.5 & 18.3 	& 20.9\\ 
	                & P     & 17.5 	& 15.9 &  14.8	& 14.1 & 13.5 	& 13.1 & 11.8 	& 10.4\\
& $\textbf{E}$   &  $\textbf{-8.6}$ &  $\textbf{-0.8}$ & $\textbf{+1.0}$ & $\textbf{+8.0}$ &$\textbf{+14.2}$ & $\textbf{+16.4}$ & $\textbf{+6.5}$ & $\textbf{+10.5}$\\
\hline
G035.21$-1.74$ & O    & 11.0 & 15.1 & 19.6 & 23.6  & - & 26.3 & 15. 6 & 22.1\\
			& P     & 18.5 & 17.1 & 16.0 & 14.9 & - & 14.5 & 12.7 & 11.2\\ 
& $\textbf{E}$  & $\textbf{-7.5}$ & $\textbf{-2.0}$ & $\textbf{+3.6}$ & $\textbf{+8.7}$ & - & $\textbf{+11.8}$ & $\textbf{+2.9}$ & $\textbf{+10.9}$\\
\hline
G035.57$-0.01$ & O   & 30.8 & 25.6 & 14.1 & 14.6 & 25.1 & 24.6 & 15.0 & 22.4\\ 
			&  P   & 19.3 & 17.8 & 16.6 & 15.6 & 15.1 & 14.6 & 13.2 & 11.6\\ 
& $\textbf{E}$   & $\textbf{+11.5}$ & $\textbf{+7.8}$ & $\textbf{-2.5}$ & $\textbf{-1.0}$ & $\textbf{+10.0}$ & $\textbf{+10.0}$ & $\textbf{+1.8}$ & $\textbf{+10.7}$\\
\hline
G040.53+2.53 & O    & 10.4 & 12.1 & 7.2  & 12.8 & - & 6.4 & 9.0 & 8.8\\
		     &  P    &   9.8 & 9.0   & 8.4  & 7.9   & - & 7.4 & 6.7 & 5.9\\ 
& $\textbf{E}$  & $\textbf{+0.7}$ & $\textbf{+3.1}$ & $\textbf{-1.2}$ & $\textbf{+4.9}$ & - & $\textbf{-1.0}$ & $\textbf{+2.3}$ & $\textbf{+2.9}$\\
\hline
G043.17$-0.01$ & O    & 89.2  & 84.0  & 57.1  & 67.7  & 89.8  & 89.2  & 54.00 & 75.5\\   
(W49)	  	& P     & 74.8  & 69.2  & 64.5  & 59.5  & 57.8  & 55.7  & 51.2 & 45.2\\
& $\textbf{E}$  & $\textbf{+14.4}$ & $\textbf{+14.8}$ & $\textbf{-7.4}$ & $\textbf{+8.2}$ & $\textbf{+32.0}$ & $\textbf{+33.5}$ & $\textbf{+2.8}$ & $\textbf{+30.3}$\\
\hline
G045.12+0.15 & O     & 4.5  & 6.8  & 5.3  & 11.2 & 11.0 & 11.0  & 9.4  & 9.8\\ 
	 	      & P     & 6.6  & 6.1  & 5.7  & 5.4  & 5.2  & 5.0  & 4.6  & 4.0\\
& $\textbf{E}$ & $\textbf{-2.1}$ & $\textbf{+0.7}$ & $\textbf{-0.4}$ & $\textbf{+5.8}$ & $\textbf{+5.8}$ & $\textbf{+6.0}$ & $\textbf{+4.8}$ & $\textbf{+5.8}$\\
\hline
G045.48+0.08 & O   & 18.6 & 23.5 & 16.3 & 30.4  & 21.4 & 27.9  & 16.3 & 15.5\\  
		     &  P    & 21.2 & 19.6 & 18.3 & 17.1  & 16.7 & 16.1  & 14.5 & 12.8\\
& $\textbf{E}$ & $\textbf{-2.6}$ & $\textbf{+3.9}$ & $\textbf{-2.0}$ & $\textbf{+13.3}$ & $\textbf{+4.7}$ & $\textbf{+11.8}$ & $\textbf{+1.8}$ & $\textbf{+2.7}$\\
\hline
\hline
Mean of $\frac{\textbf{O}}{\textbf{P}}$ & & 0.90$\pm$0.11 & 1.13$\pm$0.06 & 0.96$\pm$0.05  & 1.52$\pm$0.11 & 1.65$\pm$0.14 & 1.75$\pm$0.13 & 1.41$\pm$0.10 &
1.91$\pm$0.23 \\
Mean of $\frac{\textbf{E}}{\textbf{P}}$ & & -0.10$\pm$0.11& 0.09$\pm$0.06 & -0.04$\pm$0.05 & 0.52$\pm$0.11 & 0.65$\pm$0.14 & 0.75$\pm$0.13 & 0.41$\pm$0.10 &
0.91$\pm$0.23\\
\hline
\end{tabular}
\label{fitted_measured}
\end{center}
\end{table*}
\end{landscape}
\normalsize

\section{Anomalous emission in the selected HII regions}
\label{sec:Anomalous emission in the selected HII regions}

In this section we discuss the properties of the excess emission from the warm dust associated with  9 selected HII regions. These 
include some of the brightest HII regions in the Galaxy. We quantify the anomalous emission process as a function of 
radio frequency and of the FIR properties of the dust.

\subsection{Excess emission at radio frequencies}
\label{sec:Excess emission at radio frequencies}

\begin{figure}
\begin{center}
\includegraphics[width=8.5cm]{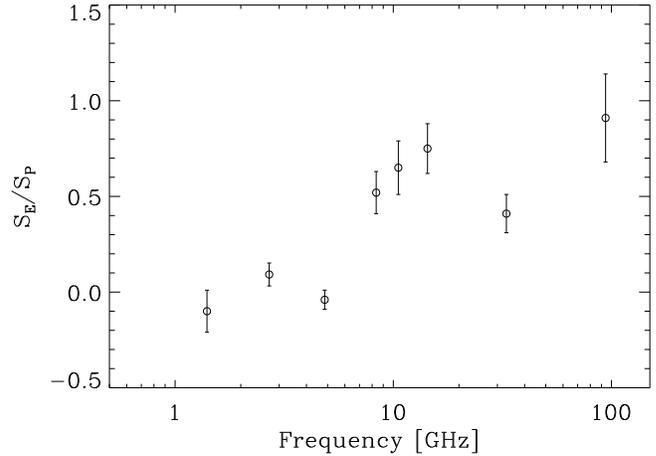}
\caption{The average ratios of the  excess emission, $S_{E}$, to the predicted free-free flux density, $S_{P}$, as a function of frequency. $1 \sigma$ errors are shown.}
\label{excess-to-free-free}
\end{center}
\end{figure}

The excess emission above the free-free emission for each HII region is defined and discussed in Sections 
~\ref{sec:Spectra of individual HII regions} and ~\ref{sec:Determining the free-free component}. Basically the 2.7- and 4.85- 
GHz data define the HII spectrum, with confirmation by the 1.4~GHz data. Values of the predicted (P) free-free and the excess (E) 
flux densities are given in Table~\ref{fitted_measured} for each of the 9 HII regions at each of the 8 spectral frequencies.
The mean value of the ratio of the excess to the HII flux density (E/P) and its r.m.s. are given in Table~\ref{fitted_measured} 
and plotted for the 8 frequencies in Fig.~\ref{excess-to-free-free}.

The excess emission at frequencies in the range 8.35 to 94 GHz is clearly seen. At individual frequencies there is a scatter between the 
HII regions of $\sim 25$~per cent about the mean; this will include a measurement error and will have a component due to 
differences in the emission process in HII regions. The 33 GHz excess relative to the HII emission has the smallest 
scatter. At this frequency the excess is $41\pm10$~per cent of the free-free emission. It is interesting to note that the bright 
diffuse emission on the Galactic plane has a similar ratio of anomalous to free-free emission (Kogut et al. 2009; Alves et al. 2010).

The 94 GHz excess relative to the free-free has the highest value. As we shall demonstrate below, this excess is due to thermal 
(vibrational) emission from the larger dust grains. The spinning dust spectrum is not expected to extend up to such 
high frequencies or at least will be small at 94 GHz (Ali--Ha{\"i}moud et al. 2009).

\subsection{33 GHz excess compared with FIR bands}
\label{sec:33 GHz excess compared with FIR bands}

We now compare the 33-GHz data with the IRAS FIR bands since there is a substantial body of data (ground-based and WMAP), which should provide a comprehensive 
picture of anomalous emission properties at a frequency near the centre of  the emission spectrum of spinning dust.

As we shall see below (Section 8.3), there is a strong correlation between the 4 IRAS bands. However, they represent 
different types of dust; the 60- and 100-$\mu$m bands are from larger grains, while the 25- and 12-$\mu$m bands 
are from very small grains (VSGs) and polycyclic aromatic hydrocarbons (PAHs). The spinning dust emission is from the latter grains.

Table~\ref{excess33_to_thermal} lists the flux densities for the 9 HII regions in the 33 GHz and the four IRAS bands. The 33 GHz excess is also 
given as a ratio to the IRAS flux densities. The mean values and the errors of these ratios for 
the 4 IRAS bands are listed. It can be seen that there is a marginally smaller scatter  of the 33 GHz excess with the 
25- and 12-$\mu$m bands ($6.0\sigma$ and $5.8\sigma$) than with the 100- and 60-$\mu$m bands ($5.4\sigma$ and $5.0\sigma$). Such a 
difference would be expected in a spinning dust scenario.

\begin{figure}
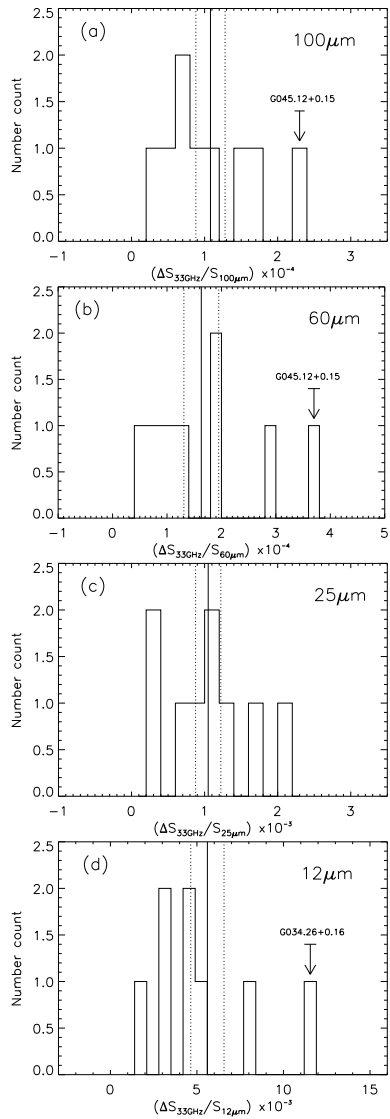

\begin{center}
\includegraphics[width=5cm]{excess-to-100m-histo.epsi}
\includegraphics[width=5cm]{excess-to-60m-histo.epsi}
\includegraphics[width=5cm]{excess-to-25m-histo.epsi}
\includegraphics[width=5cm]{excess-to-12m-histo.epsi}
\caption[]{Histograms (a), (b), (c) and (d) showing the spread in the ratio of  the excess 
33 GHz integrated flux density to those of the IRAS bands at 100, 60, 25 and 12 $\mu$m, respectively. The vertical  solid and the dotted lines show 
the mean and the 1 $\sigma$ values respectively.}
\label{IRAS_histograms_excess}
\end{center}
\end{figure}

Fig.~\ref{IRAS_histograms_excess} shows this scatter in  the distribution of 33 GHz excess relative to the FIR emission in the 4 IRAS bands. 
It is likely that the bulk 
of the scatter is due to the 33-GHz data; the radio and FIR emission is co-located as indicated in the positions of the individual HII regions 
(Section~\ref{sec:Radio positions relative to the dust}). For example G045.12+0.15 is high relative to the rest of the sample. This could be 
due to systematics in the 33 GHz data or 
differences in intrinsic spinning dust emission at 60 and 100 $\mu$m.

\subsection{The anomalous radio spectrum relative to the 100 micron dust emission}

The ratio of the radio emission relative to FIR dust emission expressed as a flux density or surface brightness is commonly  
used as an indicator of the strength of the anomalous emission. We now use our radio data covering 8 frequencies in the range 
1.4 to 94~GHz to derive a spectrum for the dust-correlated radio emission relative to the 100~$\mu$m emission. The ratio of excess 
radio to total 100~$\mu$m flux density in each of the 8 frequency bands is given for the 9 HII regions in Table~\ref{spinning_to_thermal}.

\begin{figure}
\begin{center}
 \includegraphics[width=6.0cm]{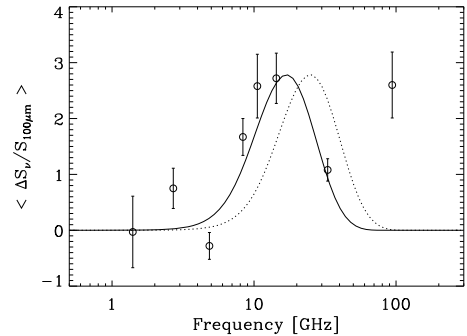}
\caption[]{The  spectrum of the averaged radio emission relative to 100~$\mu$m emission for the 9 HII regions.  The values are in units of $10^{-4}$ and the plotted errors are $1\sigma$. 
The dotted curve is the DL98 model for rotational emission from a spinning dust grains in warm ionized medium (WIM). The solid line is 
the same model shifted by a factor of $0.68$ to lower frequencies.}
\label{spinning_fit}
\end{center}
\end{figure}

Table~\ref{spinning_to_thermal} also lists the radio emissivity relative to 100~$\mu$m at each frequency, averaged across the 
9 HII regions. Fig.~\ref{spinning_fit} shows the resulting spectrum of excess emissivity in the frequency range 1.4 to 94 GHz. We will 
argue below that the 94~GHz is mostly thermal (vibrational) emission from large grains rather than spinning dust emission.

Fig.~\ref{spinning_fit} illustrates the central result of this investigation. It defines the spectrum of warm dust in 
HII regions which can be compared with that of cool dust clouds (Watson et al., 2005; Casassus et al., 2006). 
The figure also shows the Draine \& Lazarian (1998) model for the warm ionized
 medium (WIM). Our data 
would give a better fit if the WIM spectrum were moved to frequencies lower by a
factor of $0.68$. A wide 
range of spectral shapes are predicted by Ali-Ha{\"i}moud et al. (2009), some of which give a better fit to the data. A shift in the position of the peak could also indicate different chemical compositions (Iglesias-Groth, 2005). We note that the lowest peak frequencies observed theoretically, are $\sim 15$~GHz (Ali-Ha{\"i}moud et al. 2009), similar to what is observed in Fig.~7. Also, the WIM model of Ysard \& Verstraete (2010) had the lowest peak frequency ($\sim 20$~GHz) of the models representing typical ISM environments.

\subsection{Comparison of the present VSA results with other 30-GHz measurements}
\label{sec:Comparison of the present VSA results with other 30 GHz measurements}

We first consider the anomalous emission from HII regions. A summary of the published data is given in 
Table~\ref{current_emissivities}. These include the CBI results (Dickinson et al., 2006, 2007,  2009) 
at 31 GHz. These have been converted to emissivities in $\mu$K (MJy/sr)$^{-1}$ at 33 GHz for comparison 
with the present data. In the case of RCW175 the results for its two components are given separately.
We note that the VSA flux density (7.8~Jy) is somewhat greater than that given by the CBI (5.97~Jy). 
This probably results from the fact that the CBI has resolved out some of the extended emission.

There is clearly good consistency between the results for the warm dust clouds associated with HII regions in 
the various investigations. The weighted average emissivity is $4.65\pm0.40$ (11.6 $\sigma$) $\mu$K (MJy/sr)$^{-1}$ 
at 33 GHz.

It is of particular interest to compare the radio emissivity of the warm ($\sim 40$ K) dust in HII regions with that 
in the general field where dust temperatures are typically $\sim 20$ K. The data, corrected to 33 GHz where 
necessary, are given in Table~\ref{cooldust_emissivities}.

\begin{landscape}
\begin{table}
\begin{center}
\fontsize{7}{10pt}\selectfont
\renewcommand{\arraystretch}{1.}		
\caption[]{The comparison 
of the 33 GHz excess for the 9 HII regions relative to the integrated flux densities in the four IRAS bands. Column 1 lists 
the sources,  columns 2 and 3 list the observed ($S_{33}$) and the excess ($\Delta S_{33}$) 
 flux densities at 33 GHz. The observed  flux densities in the IRAS bands 
are shown in columns 4-7. Columns 8-12 list the ratios of the excess  flux densities at 
33 GHz to the flux densities at 100, 60, 25  and 12$\mu$m. The mean ratios of 33 GHz excess relative to the IRAS flux densities is given in heavy type.}
\vspace{0.5cm}
\begin{tabular}{|l|rr|rrrr|rrrr|}
\toprule
 & $S_{33}$ &  $\Delta S_{33}$ & $S_{100}$ & $S_{60}$ & $S_{25}$ & $S_{12}$ & $\frac{\Delta S_{33}}{S_{100}}$ & $\frac{\Delta S_{33}}{S_{60}}$ & $\frac{\Delta S_{33}}{S_{25}}$ & $\frac{\Delta
S_{33}}{S_{12}}$ \\
Source & [Jy] & [Jy] & $\times10^{4}$ [Jy] &$\times10^{4}$ [Jy] & $\times10^{2}$ [Jy] & $\times10^{2}$  [Jy] & $\times10^{-4}$ &$\times10^{-4}$ &  $\times10^{-4}$ & $\times10^{-4}$\\
\hline
1 & 2 & 3 & 4 & 5 & 6 & 7 & 8 & 9 & 10 & 11\\
\midrule
G029.02$-$0.55  	&  7.8 $\pm$1.6 	&  +3.1$\pm$1.7  	& 1.8$\pm$0.4 		& 1.1$\pm$0.2  		& 9.1$\pm$2.0  		& 7.0$\pm$1.4	&  1.72  &  2.82  & 20.2  & 44.3\\

G029.92$-$0.05 	& 15.9$\pm$3.2  	&  +5.4$\pm$3.5  	& 3.8$\pm$0.8  		& 2.9$\pm$0.6  		& 49.1$\pm$9.8   	& 7.4$\pm$1.5   &  1.42  & 1.86   & 11.0  & 73.3\\
				
G034.26+0.16 		& 18.3$\pm$3.7  	&  +6.5$\pm$4.0  	& 5.5$\pm$1.1 		& 3.5$\pm$0.7  		& 39.4$\pm$7.9   	& 5.5$\pm$1.1   &  1.18  & 1.86   & 16.5   & 118.4\\

G035.21$-$1.74 	& 15.6$\pm$3.1 	&  +2.9$\pm$3.6  	& 3.7$\pm$0.7  		& 3.0$\pm$0.6  		& 46.3$\pm$9.3   	& 9.8$\pm$2.0   &  0.78  &  0.97  &  6.3   & 29.6\\

G035.57$-$0.01 	& 15.0$\pm$3.0 	& +1.8$\pm$3.6  	& 2.7$\pm$0.5  		& 1.7$\pm$0.3  		& 15.8$\pm$3.2   	& 3.3$\pm$0.7	&  0.67  &1.06    & 11.4   & 54.9\\

G040.53+2.53 		&  9.0$\pm$1.8  	&  +2.3$\pm$3.0  	& 2.6$\pm$0.5  		& 1.8$\pm$0.4  		& 25.2$\pm$5.0  	& 4.8$\pm$1.0   &  0.88  & 1.28   & 9.1   & 47.6\\
	
G043.17$-$0.01		& 54.0$\pm$10.8 	&  +2.8$\pm$1.2  	& 6.8$\pm$1.4  		& 4.5$\pm$0.9 		& 75.0$\pm$15.0   	& 8.2$\pm$1.6    &  0.41 &  0.62  &  3.7   & 34.2\\
		
G045.12+0.15 		&  9.4$\pm$2.0  	&  +4.8$\pm$2.0  	& 2.1$\pm$0.4  		& 1.3$\pm$0.3  	& 38.9$\pm$7.8   	& 5.9$\pm$1.2   &  2.29  &  3.63  & 12.3   & 81.6\\
		
G045.48+0.08 		& 16.3$\pm$3.3  	&  +1.8$\pm$3.9  	& 4.8$\pm$1.0  		& 3.1$\pm$0.6  		& 45.6$\pm$9.1   	& 8.8$\pm$1.8   &  0.38  &  0.58  &  3.9   & 20.6\\

\hline 
\textbf{Mean}   &  &  & & & & &    \textbf{1.08}$\pm$\textbf{0.20}	&  \textbf{1.63}$\pm$\textbf{0.32}  & 
\textbf{10.5}$\pm$\textbf{1.7} 	&  \textbf{56.1}$\pm$\textbf{ 9.6}\\
\textbf{$\sigma$}     &  &    & & & & &     \textbf{5.4$\sigma$} &  \textbf{5.0$\sigma$}  &  \textbf{6.0$\sigma$}&
\textbf{5.8$\sigma$}\\
\bottomrule
\end{tabular}
\label{excess33_to_thermal}
\end{center}
\end{table}
\end{landscape}
\normalsize

The all-sky WMAP analysis with the Kp2 mask refers to dust at intermediate 
and high Galactic latitudes (Davies et al., 2006) where the 15 regions included the brighter FIR dust clouds at 
intermediate latitudes on scales of $1^{\circ}-5^{\circ}$. 
The more compact regions G159.6$-$18.5 and 
LDN 1622 each contain fine structure and relatively weak associated HII regions; they have the highest emissivity.

In summary, the 33 GHz emissivity of HII region dust clouds is $\sim40$~per cent of that of dust in the diffuse and 
extended ISM and is $\sim20$~per cent of that in more compact dust clouds. We will discuss possible reasons for 
these different emissivities in Section~\ref{sec:Conclusions}.

\section{Other physical properties of the 9 HII regions}
\label{sec:Other physical properties of the 9 HII regions}

We have further data at the 8 radio frequencies and 4 FIR frequencies which can be used to derive additional physical 
properties of the 9 HII regions - positions, diameters, SEDs and dust temperatures.

\subsection{Radio positions relative to the dust}
\label{sec:Radio positions relative to the dust}

We use the 2.7 GHz data to define the position of the free-free emission and the 100 $\mu$m IRAS data for the dust.
In order to directly intercompare the 33 GHz data, the 13~arcmin smoothed positions at 2.7 GHz and 100 ~$\mu$m
are used since a number of HII regions have complex structures. Fig.~\ref{position_offsets} shows the relative 
positions at 2.7 GHz, 33 GHz and 100 $\mu$m for the 9 HII regions.

The mean of the offsets, excluding RCW175, are shown in
Fig.~\ref{position_offsets} as dashed lines. The offset between 33~GHz 
and 100~$\mu$m positions is $(\Delta \ell, \Delta \emph{b}) =(-0.13, -0.33)$~arcmin. The offsets from 2.7~GHz are larger; the offset from 33~GHz is 
$(\Delta \ell, \Delta \emph{b}) = (1.18, -0.34)$~arcmin and from 100~$\mu$m is $(\Delta \ell, \Delta \emph{b}) = (-1.44, -0.05)$~arcmin. The~2.7 GHz 
positions appear to be displaced by $\sim 1.2$ arcmin from those at 33~GHz and 100~$\mu$m. There is no obvious 
reason for this displacement; the 2.7~GHz positions are accurate to $20$~arcsec in $\ell$ and $33$~arcsec in $\it{b}$ (Reich et al., 1990b). 

The scatter about the mean offset is similar for each of the 3 comparisons with an r.m.s. of $1.5$~arcmin, excluding RCW175 (triangle). 
This scatter results from the noise on the 3 maps. 

These results indicate that the free-free and the dust components agree in position to $\pm 1$~arcmin when the 2.7 GHz offset 
is allowed for.  The tight correlation between the 33 GHz and 100 $\mu$m confirms this result since $\sim 70$~per cent of the 33~GHz emission is due to free-free and $\sim 30$~per cent is from anomalous emission (see Table~\ref{fitted_measured}).

\subsection{Radio diameters relative to the dust}
\label{sec:Radio diameters relative to the dust}

The 2.7 GHz, 33 GHz and 100 $\mu$m data, at the $13$~arcmin resolution of the VSA beam, can be used to intercompare 
the diameters of the free-free and dust emission in the 9 HII regions. Fig.~\ref{diameters} shows this intercomparison. 
The diameters are determined for Gaussian fits to the emission. They are the intrinsic diameters derived from the beam 
broadening  observed. Diameters less than $3$ arcmin will be uncertain; they broaden the beam by $\leq 0.4$ arcmin.

\begin{table*}
\begin{center}
\fontsize{6}{10pt}\selectfont
\renewcommand{\arraystretch}{1.}
\caption[]{The comparison of the excess radio flux density relative to 100 $\mu$m from 1.4 to 94~GHz for the 9 HII regions.  
All ratios are given in units of $\times10^{-4}$.}
\vspace{0.5cm}
\begin{tabular}{|l r r r r| r r r r|}
\toprule
Source& $\frac{\Delta S_{1.4}}{S_{100\mu}}$ & $\frac{\Delta S_{2.7}}{S_{100\mu}}$ & $\frac{\Delta S_{4.85}}{S_{100\mu}}$ & $\frac{\Delta S_{8.35}}{S_{100\mu}}$  &  $\frac{\Delta
S_{10.55}}{S_{100\mu}}$ &  $\frac{\Delta S_{14.35}}{S_{100\mu}}$ & $\frac{\Delta S_{33}}{S_{100\mu}}$  & $\frac{\Delta S_{94}}{S_{100\mu}}$  \\
\midrule
\midrule
G029.02$-$0.55		& $-0.17$	& 0.00      & -          	&   2.28      & -        & 2.33      & 1.72   & 5.67\\
G029.92$-$0.05 	& $-1.66$  & 0.00      & -          	&  0.74       &  0.76  & 2.42      & 1.42   & 0.03\\
G034.26$+$0.16	& $-1.56$  & $-0.15$ & 0.18 	&  1.45      & 2.58    & 2.98      & 1.18   & 1.91\\
G035.21$-$1.74		& $-2.03$  & $-0.54$ & 0.97      	&  2.35      & -         & 3.19      & 0.78   & 2.95 \\
G035.57$-$0.01 	& 4.26      & 2.89      & $-0.93$  &  $-0.37$  & 3.70    & 3.70      & 0.67   & 3.96 \\
G040.53$+$2.53 	& 0.27      & 1.19      & $-0.46$  &  1.88      & -         & $-0.38$  & 0.88    & 1.12 \\
G043.17$-$0.01         & 2.12      & 2.18      & $-1.09$  &  1.21      & 4.71    & 4.93      & 0.41    & 4.46 \\
G045.12$+$0.15	& $-1.00$ 	& 0.33     & $-0.19$  &  2.76      & 2.76    & 2.86       & 2.29    & 2.76 \\
G045.48$+$0.08 	& $-0.54$ 	& 0.81     & $-0.42$  &  2.77      & 0.98    & 2.46       & 0.38    & 0.56 \\
\hline
Mean 			& 0.31$\pm$0.63 	& 0.75$\pm$0.36	& -0.28$\pm$0.24 	&  1.67$\pm$0.33  	&  2.58$\pm$0.57  	& 2.72$\pm$0.45 	&
1.08$\pm$0.20	&2.60$\pm$0.59	\\

$\sigma$    		& 0.5$\sigma$		& 2.1$\sigma$		& 1.1$\sigma$		& 5.1$\sigma$ 		& 4.2$\sigma$ 		& 6.1$\sigma$ 		 &
5.4$\sigma$ 	&4.3$\sigma$\\
\bottomrule
\end{tabular}
\label{spinning_to_thermal}
\end{center}
\end{table*}

\normalsize

\begin{table*}
\begin{center}
\fontsize{7}{10pt}\selectfont
\caption[]{Comparison of 33 GHz radio emissivities, relative to $100~\mu$m, of dust in HII regions.}
\begin{tabular}{l c c l}
\toprule
HII regions & Instrument & Dust emissivity (33 GHz) & Reference\\
& & $\mu$K (MJy/sr)$^{-1}$ & \\
\hline
6 Southern HII regions       	&  CBI  & $2.9\pm1.5$  		 		& Dickinson et al.(2007)\\
LPH96                            	&  CBI  & $5.2\pm2.1$   				& Dickinson et al.(2006)\\
G029.01$-$0.6			&  CBI  & $4.7\pm0.9$   				& Dickinson et al.(2009)\\
G029.01$-$0.7		       	&  CBI  & $6.8\pm1.1$  		 		& Dickinson et al.(2009)\\
9 Northern HII regions        &  VSA & $3.9\pm0.8$   				& This work\\ 
\hline
Weighted average		& 	    & 	$4.65\pm0.40$ (11.5$\sigma$)	&\\
\bottomrule
\end{tabular}
\label{current_emissivities}
\end{center}
\end{table*}
\normalsize

\begin{table*}
\begin{center}
\fontsize{7}{10pt}\selectfont
\caption[]{Comparison of 33 GHz emissivities relative to 100 $\mu$m in cool (field) and warm (HII region) dust clouds.}
\begin{tabular}{l c c l l}
\hline
Region 	& Instrument   		& Dust emissivity (33 GHz)  		& Morphology					&  Reference\\
                 		&              		& $\mu$K (MJy/sr)$^{-1}$     		& Size       					& \\
\hline
G159.6-18.5      		& COSMOSOMAS    	&  $17.8\pm0.3$             		 	&  $1^{\circ}.6\times1^{\circ}.0$  	 	&  Watson et al. (2005)\\
LDN1622          		& CBI          		&  $21.3\pm0.6$              			&  $<10$~arcmin    					&  Casassus et al. (2006)\\
15 WMAP regions 	& WMAP         		&  $11.2\pm1.5$              			&   $1^{\circ} - 5^{\circ}$			&  Davies et al. (2006)\\
All-sky WMAP (Kp2)  & WMAP        	 	&  $10.9\pm1.1$              			&  Diffuse   					&  Davies et al. (2006)\\
\hline
HII regions (mean)	&       			&  $4.65\pm0.40$ (11.5$\sigma$)  	&  							& This work\\
\hline
\end{tabular}
\label{cooldust_emissivities}
\end{center}
\end{table*}
\normalsize

\begin{figure*}
\begin{center}
\includegraphics[width=7.5cm]{pos_33_100.epsi}
\includegraphics[width=7.5cm]{pos_33_27.epsi}
\includegraphics[width=7.5cm]{pos_27_100.epsi}
\caption[]
{Relative position offsets of the 9 HII regions at 2.7 GHz, 33 GHz and 100 $\mu$m dust.  (a) the 100 $\mu$m relative to 33 GHz. (b) 
2.7 GHz relative to 33 GHz (c) 2.7 GHz relative to 100 $\mu$m. The the dashed
lines show the mean offset in each plot, excluding RCW175.}
\label{position_offsets}
\end{center}
\end{figure*}

\begin{figure*}
\begin{center}
\includegraphics[width=7.5cm]{dims_33to100.epsi}
\includegraphics[width=7.5cm]{dims_33to25.epsi}
\includegraphics[width=7.5cm]{dims_33to27.epsi}
\includegraphics[width=7.5cm]{dims_27to100.epsi}
\caption[]
{Comparison of the deconvolved 33 GHz dimensions of HII regions to those at (a) 100 $\mu$m, (b) 25 $\mu$m  and (c) 2.7 GHz . 
The 2.7 GHz (free-free)  dimensions are compared to those of dust (100 $\mu$m) in (d). Major and minor axes diameters are given in 
each plot.}
\label{diameters}
\end{center}
\end{figure*}

The intercomparison of the 2.7 GHz, 33 GHz and the 100 $\mu$m data shows close similarity of diameters at 
these 3 wavelengths. The tight correlation between the 2.7 GHz (free-free) and the 100 $\mu$m (dust) diameters 
emphasizes the close agreement in the spatial distribution of these 2 ISM components.  

The agreement between the free-free and dust positions along with the similarity of the diameters shows that ionized gas 
and warm dust are well mixed in HII regions. This is not the situation in the general ISM as was found in the 15-region investigation 
(Davies et al., 2006).

\subsection{The radio - FIR spectrum of warm thermal dust in HII regions}
\label{sec:The radio - FIR spectrum of warm thermal dust in HII regions}

We now determine the SED of thermal dust in HII regions extending from 94 GHz, where the excess due to dust 
is unambiguously detected at radio frequencies, to the IRAS FIR frequencies. Table~\ref{radio_FIR_SED_table} 
shows the mean flux density ratio of the 4 bands relative to the 100 $\mu$m values for the 9 HII regions. 

\begin{table}
\begin{center}
\fontsize{8}{10pt}\selectfont
\renewcommand{\arraystretch}{1.}
\caption[The comparison of mean dust emissivities.]{The summary table of the mean emissivities in the IRAS bands and the 94 GHz 
excess emission with respect to 100 $\mu$m thermal dust. Note that the 94~GHz value should be multiplied by $10^{-4}$.}
\label{radio_FIR_SED_table}
\begin{tabular}{lcccc}
\toprule
 & $\frac{\Delta S_{94GHz}}{S_{100\mu\rm{m}}}$  & $\frac{S_{60\mu\rm{m}}}{S_{100\mu\rm{m}}}$ & $\frac{S_{25\mu\rm{m}}}{S_{100\mu\rm{m}}}$ & $\frac{S_{12\mu\rm{m}}}{S_{100\mu\rm{m}}}$ \\
\midrule
Mean  & $2.60$  & $0.67$ & 	 $0.103$ & $0.0203$\\ 
$\sigma$ & $4.4\sigma$   & $30.2\sigma$ & $7.9\sigma$  & $7.1\sigma$ \\

\bottomrule
\end{tabular}
\end{center}
\end{table}
\normalsize

It is clear that the 60~$\mu$m and 100~$\mu$m values are tightly correlated (at $30\sigma$) while the 25~$\mu$m 
and 12~$\mu$m are less correlated with 100~$\mu$m (at $7.9\sigma$ and $7.1\sigma$, respectively). The 94~GHz dust emission shows a 
significant ($4.4\sigma$) detection of the long wavelength dust emission. The scatter may be due to variation 
in the emissivity index $\beta$ between the 9 dust clouds.

\begin{figure}
\begin{center}
\includegraphics[width=7.0cm]{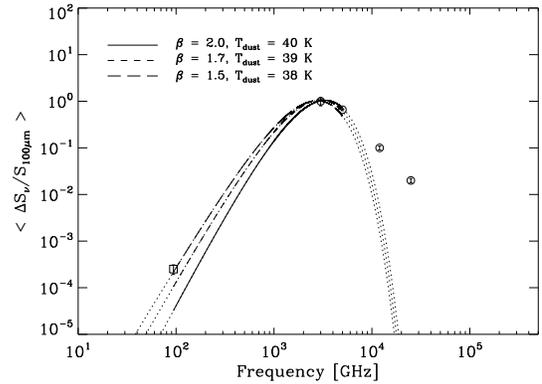}
\caption[]{The mean emissivities of the 9 HII regions in the IRAS bands and the 
94 GHz excess emission with respect to 100~$\mu$m thermal dust. 
The $\beta=$1.5, 1.7 and 2.0 spectra for thermal dust are shown separately. The best fit is for a dust temperature of 39 K and 
$\beta=1.5$.}
\label{fitex}
\end{center}
\end{figure}

Fig.~\ref{fitex} is the plot of the SED relative to 100~$\mu$m of the warm dust in the 9 HII regions of the present 
study. Also plotted are spectra through the 60 and 100~$\mu$m points for assumed dust emissivity indices  
of $\beta=$1.5, 1.7 and 2.0. The best fitting dust temperature, which includes the 94~GHz value, is 39~K for $\beta=+1.5$. 
The very small grain dust emission represented by the 
12 $\mu$m and 25 $\mu$m points shows as an excess above the extrapolated large grain dust emission.

\subsection{Temperature dependence of relevant parameters}
\label{sec:Temperature dependance of relevant parameters} 

The IRAS data can also be used to derive the dust temperature of each HII region. Kuiper et al. (1987) give recipes for 
deriving the temperature of the large grain component in the normal ISM environment, using the 60 and 100~$\mu$m 
data. This treatment assumes a value of $\beta$ and does not take account of the 94~GHz in determining $\beta$. 
It is a complex process to determine an accurate dust temperature (e.g. Dupac et al.~2003), so accordingly we will adopt the 
ratio, R, of 60~$\mu$m to 100~$\mu$m flux density as a temperature indicator in our investigation of dust properties. 
A higher value of R represents a higher temperature. The 
parameters considered are the anomalous emission ($\Delta S_{33 \textrm{GHz}}/ S_{100\mu\textrm{m}}$), the thermal 
dust ($\Delta S_{94 \rm{GHz}}/S_{100\mu\rm{m}}$) and the free-free emission ($S^{ff}_{33 \rm{GHz}}/S_{100\mu\rm{m}}$).
The values of each parameter are listed in Table~\ref{temperatures}.

\begin{table*}
\begin{center}
\fontsize{7}{10pt}\selectfont
\renewcommand{\arraystretch}{1.0}
\caption[]{The parameters used for investigating
temperature dependences   in the 9 HII regions. R, the ratio between the 60 $\mu$m and the 100 $\mu$m flux densities, is the temperature 
indicator.}
\label{temperatures}
\begin{tabular}{l c c c c c}
\toprule
 & R & $\frac{\Delta S_{33\rm{GHz}}}{S_{100\mu\rm{m}}}$ & $\frac{S^{ff}_{33\rm{GHz}}}{S_{100\mu\rm{m}}}$ & $\frac{\Delta S_{94\rm{GHz}}}{S_{100\mu\rm{m}}}$ & 
$\frac{S_{25\mu\rm{m}}}{S_{100\mu\rm{m}}}$ \\
Source &  &            $\times10^{-4}$ 			 &           $\times10^{-4}$ 		&      $\times10^{-4}$	  &      $\times10^{-4}$ \\
\midrule
G029.02$-$0.55      	 	&  0.60 & 1.72 & 2.61 & 5.67 & 0.051\\
G029.92$-$0.05 		&  0.76 & 1.42 & 2.76 & 0.03 & 0.129\\
G034.26+0.16 			&  0.64 & 1.18 & 2.14 & 1.91 & 0.072\\
G035.21$-$1.74 		&  0.81 & 0.78 & 3.43 & 2.95 & 0.125\\
G035.57$-$0.01 		&  0.63 & 0.67 & 4.89 & 3.96 & 0.059\\
G040.53+2.53 			&  0.69 & 0.88 & 2.58 & 1.12 & 0.097\\
G043.17$-$0.01          	&  0.66 & 0.41 & 7.53 & 4.46 & 0.110\\
G045.12+0.15 			&  0.62 & 2.29 & 2.19 & 2.76 & 0.185\\
G045.48+0.08 			&  0.65 & 0.38 & 3.02 & 0.56 & 0.095\\
\hline
Mean				&  & $1.08\pm0.20$ & $3.46\pm0.55$ & $2.60\pm0.59$ & $0.103\pm0.0131$ \\
$\sigma$ 				&  & $5.4\sigma$ & $6.3\sigma$ & $4.4\sigma$ & $7.9\sigma$ \\
\bottomrule
\end{tabular}
\end{center}
\end{table*}
\normalsize

\begin{figure}
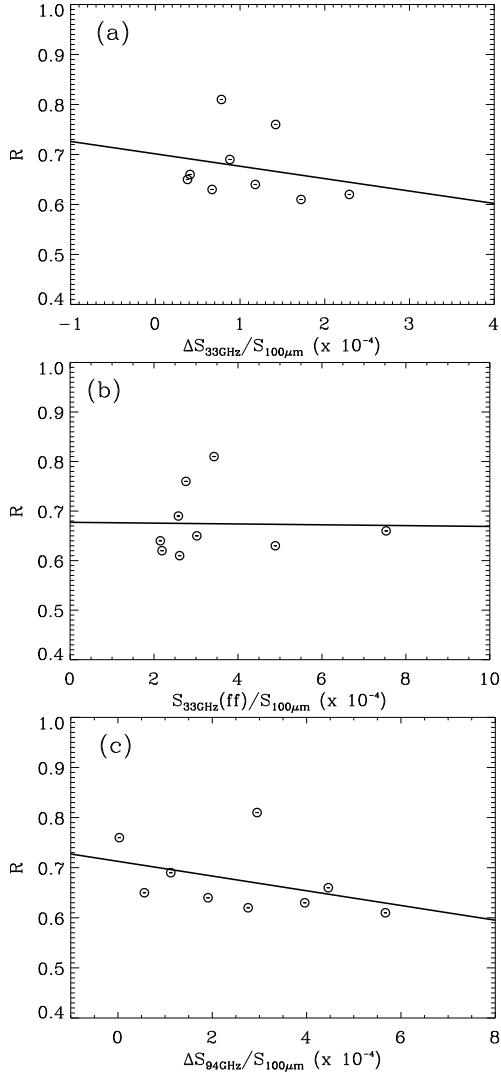

\begin{center}
\includegraphics[width=6.5cm]{temp-dep-33excess-to-100m.epsi}	
\includegraphics[width=6.5cm]{temp-dep-free-free-to-100m.epsi}	
\includegraphics[width=6.5cm]{temp-dep-94GHz-to-100m.epsi}	
\caption[]
{The temperature dependence of (a) anomalous emission, (b)  free-free emission and (c) 94 GHz. The temperature indicator, R, is 
the ratio of 60 $\mu$m to 100 $\mu$m flux density.}
\label{linear3-33ex-100}
\end{center}
\end{figure}

Fig.~\ref{linear3-33ex-100} (a) shows the anomalous emission for each of the 9 HII regions as a function of R, the 
ratio between the 60 and  100 $\mu$m flux densities.
This plot shows a weak correlation ($2.6  \sigma$) with a decrease in $\Delta S_{33 \textrm{GHz}}/S_{100\mu\textrm{m}}$ 
as a function of temperature. Such a decrease would be caused by a greater 100 $\mu$m emission per unit dust density 
at higher temperatures - as expected.

Fig.~\ref{linear3-33ex-100} (b) shows the free-free emission at 33 GHz relative to 100 $\mu$m as a function of 
temperature. A modest positive correlation is found suggesting the level of free-free increases 
more rapidly with temperature than the dust at 100 $\mu$m. There is no obvious physical reason for this.

Fig.~\ref{linear3-33ex-100} (c) shows the 94 GHz thermal dust emission relative to 100 $\mu$m as a function of temperature. 
This shows the strongest correlation of the study. Such a negative correlation could be due to a $\beta$ 
variation with temperature in the thermal emission spectrum and/or the increase in 100 $\mu$m emission with temperature, 
as expected for thermal dust emission.

\section{Conclusions}
\label{sec:Conclusions}

We have presented here a high sensitivity 33 GHz map made with the VSA of a $19^{\circ}$ section of the 
northern Galactic plane covering $\ell = 27^{\circ}$ to $46^{\circ}$. The VSA detects structure up to a scale of 
$\sim 50$~arcmin with a resolution of $13$~arcmin;  scales $> 50$~arcmin are resolved out. The major part of this structure detected at 33 GHz 
is free-free emission from HII regions. This data set is combined with 7 other surveys in the frequency range 
1.4 to 94~GHz and IRAS data at 100, 60, 25 and 12~$\mu$m to investigate the physical properties of the warm 
dust and ionized gas in 9 well-defined HII regions detected in the survey.

A significant outcome of this investigation is the clear detection of anomalous dust-correlated emission in 
HII regions. When combined with published data on individual HII regions the 33 GHz relative to the 100 $\mu$m 
brightness is $4.65\pm0.40$ $\mu$K  (MJy/sr)$^{-1}$ ($11.5 \sigma$). This value is 3-5 times less than that 
of cool (20 K) dust in the field. This lower radio/FIR emissivity in the warm clouds most likely arises from the 
fact that the 100 $\mu$m brightness increases as $T^{4}$ while the 33 GHz emission depends on the surface 
density of dust grains only. Further investigation is required.

The combined radio data set has been used to determine the spectrum of the excess emission in the warm 
HII region dust clouds. The mean spectrum of this excess emission relative to the 100 $\mu$m emission is 
found to peak in the region of 20 GHz. Although the spectrum is offset in frequency relative to the Draine \& 
Lazarian prediction for the warm interstellar medium, other predictions are available (Ali-Ha{\"i}moud et al. 2009) 
which may throw light on the physical conditions necessary for the anomalous emission.

The 33 GHz survey of HII regions on the Galactic plane has shown the close association between warm 
dust and free-free emission; the dust and ionized gas appear well-mixed on arcmin scales. Our survey 
has also permitted the identification of the thermal (vibrational) component of emission at 94 GHz which provides 
a measure of the dust emission at short radio wavelengths. This is an area where $\emph{Planck}$ (Tauber et al., 2010) measurements will 
provide a definitive spectrum of the radio SED of dust, both in the Galactic plane and at intermediate 
latitude. Finally, we have found only weak correlations between the dust temperature and such parameters 
as anomalous radio emissivity, free-free emission and 94~GHz emission; these correlations, although weak, probably result from the expected 
higher FIR emission per unit dust density at higher temperatures.

\section{Acknowledgments}
\label{sec:Acknowledgements}
We thank the anonymous referee for useful comments on the paper. We thank the staff of Jodrell Bank Observatory, Mullard Radio Astronomy Observatory and the Teide Observatory for assistance in the day-to-day operation of the VSA. We are very grateful to PPARC (now STFC) for the funding and support for the VSA project and the Instituto de Astrof\'{i}sica de Canarias (IAC) for supporting and maintaining the VSA in Tenerife. Partial financial support was provided by the Spanish Ministry of Science and Technology project AYA2001-1657.
CD acknowledges an STFC Advanced Fellowship and ERC grant under the FP7.  YAH thanks the King Abdulaziz City for Science and Technology for support. YAH also thanks His Highness Prince Dr Turki Bin Saud Bin Mohammad Al Saud for his personal support. JAR-M is a Ram\'on y Cajal fellow of the Spanish Ministry of Science and
Innovation (MICINN).

\bibliographystyle{mn2e}


\bsp 

\label{lastpage}

\end{document}